\documentclass[]{IEEEtran}
\usepackage{amsmath,amsfonts}
\usepackage{algorithmic}
\usepackage{algorithm}
\usepackage{array}
\usepackage{subfigure}
\usepackage{textcomp}
\usepackage{stfloats}
\usepackage{url}
\usepackage{verbatim}
\usepackage{graphicx}
\usepackage{cite}
\usepackage{chngcntr}
\usepackage{lipsum,lmodern}
\usepackage{pgf}
\usepackage[most]{tcolorbox}
\usepackage{tikz}
\usetikzlibrary{arrows,automata,shapes}
\usetikzlibrary{positioning}

\tikzset{
  state/.style={
    rectangle,
    rounded corners,
    draw=black, very thick,
    minimum height=2em,
    text width=0.6\textwidth,
    inner sep=2pt,
    text centered,
  },
}

\renewcommand\fbox{\fcolorbox{red}{white}}
\newcommand\submittedtext{%
  \footnotesize This work has been submitted to the IEEE for possible publication. Copyright may be transferred without notice, after which this version may no longer be accessible.}

\newcommand\submittednotice{%
\begin{tikzpicture}[remember picture,overlay]
\node[anchor=south,yshift=10pt] at (current page.south) {\fbox{\parbox{\dimexpr0.65\textwidth-\fboxsep-\fboxrule\relax}{\submittedtext}}};
\end{tikzpicture}%
}

\newcommand\copyrighttext{%
  \footnotesize \textcopyright \the\year{} IEEE. Personal use of this material is permitted. Permission from IEEE must be obtained for all other uses, including reprinting/republishing this material for advertising or promotional purposes, collecting new collected works for resale or redistribution to servers or lists, or reuse of any copyrighted component of this work in other works.}

\def\vOmega{\mathbf{\Omega}}
\def\tMix{t_\mathrm{mix}}

\newcommand{\sphere}{\mathbb{S}^2}
\newcommand{\br}{\mathbf{r}}
\newcommand{\bd}{\mathbf{d}}

\newcommand{\bu}{\mathbf{u}}
\newcommand{\I}{\mathrm{i}}

\begin{document}

\title{Echo Detection Using the Herglotz Wavefunction in Spatial Room Impulse Responses Measured with Spherical Microphone Arrays}

\author{Pierre Massé, Anthony Gallien, Wolfgang Kreuzer, Markus Noisternig%
\thanks{P.\ Massé, A.\ Gallien and M.\ Noisternig are from STMS, Ircam--Sorbonne Université--CNRS--Ministère de la Culture, Paris, France.}%
\thanks{W.\ Kreuzer is from the Acoustics Research Institute, Austrian Academy of Sciences, Vienna, Austria.}%
}

% The paper headers
\markboth{}%
{Massé \MakeLowercase{\textit{et al.}}: Echo Detection in Spatial Room Impulse Responses Measured with Spherical Microphone Arrays Using the Herglotz Wavefunction}

\maketitle

% Copiright notices for preprint publication at arXiv.org
\submittednotice % uncomment for submission to IEEE
% \copyrightnotice % uncomment after publication at IEEE

\begin{abstract}
 
\par Early reflections in spatial room impulse responses (SRIRs) measured using spherical microphone arrays (SMAs) play an important role in spatial audio analysis, rendering, and reverberation modeling. Their accurate localization and characterization can facilitate the analysis, processing, and manipulation of measured reverberation fields. This paper proposes a method for detecting and characterizing early reflections based on the Herglotz wavefunction formalism. The measured sound field is represented as a continuous superposition of incident plane waves, from which a localization function is derived in the spherical harmonic domain. An adaptive radial Gaussian fitting procedure is then used to estimate both the directions of arrival (DoAs) and the number of incident reflections within a single analysis frame. The proposed framework is evaluated using simulated and measured SRIRs and compared with conventional steered response power (SRP) and multiple signal classification (MUSIC) localization methods. The results demonstrate improved localization accuracy and more reliable detection of multiple simultaneous reflections, highlighting the potential of the proposed Herglotz-based formulation for the analysis of early reflections in SMA-measured SRIRs.

\end{abstract}


\section{\label{sec:intro}Introduction}
\IEEEPARstart{T}{his} paper addresses the detection and characterization of early reflections in spatial room impulse responses (SRIR) measured using spherical microphone arrays (SMA). Specifically, it introduces an approach based on the Herglotz wavefunction formalism, which represents the sound field as a superposition of plane waves equipped with a kernel function whose values specify the strength of the incident field for each direction on the sphere.

\par This work forms part of a broader research effort aimed at developing a unified space-time-frequency framework for the analysis, processing, and manipulation of SRIRs. The objective of this framework is to provide detailed analysis and modeling tools of SMA-measured SRIRs, thereby facilitating the reproduction of spatial reverberation effects. Such reproduction may be achieved either through direct multichannel convolution with measured SRIRs, or through hybrid approaches in which the late reverberation tail is synthesized by a feedback delay network (FDN)~\cite{carpentier_hybrid_2014}.

\par In direct convolution-based reproduction, the non-decaying measurement noise floor must be removed or compensated for, e.g.\ through late-reverberation resynthesis techniques~\cite{masse_robust_2020,masse_denoising_2020,masse_analysis_2022}. Beyond noise mitigation, these techniques can also be employed to modify the spatio-temporal and spectral characteristics of the late reverberation tail. Hybrid reverberators offer additional advantages, including reduced computational complexity and real-time control over reverberation decay characteristics through the manipulation of FDN parameters.

\par In both cases, a detailed ``cartography'' of the dominant early reflections significantly enhances the capabilities of the framework. Such information has considerable potential for supporting operations such as reflection filtering, redistribution, and editing, while also enabling the definition of higher-level objective or perceptual descriptors for controlling these transformations.

\IEEEpubidadjcol

\par Although the long-term objective is to construct complete early-reflection cartographies through frame-by-frame analysis of the early portion of the SRIR, the present work focuses on the first fundamental step: localizing incident reflections within a single analysis frame. Here, \textit{localization} primarily refers to the estimation of the reflections' directions of arrival (DoAs), wheras \textit{characterization} encompasses the set of attributes used to describe an individual reflection, namely its DoA, relative time of arrival (ToA), total energy, and spectral envelope.

\subsection{\label{sec:stateOfArt}Related Work}

\par The problem addressed in this paper belongs to the general class of source localization problems, for which an extensive body of literature exists covering a wide range of microphone array configurations and signal processing techniques. However, several characteristics unique to echo detection in SMA-measured SRIRs restrict the set of applicable methods in practice.

\par These limitations are discussed in further detail in Secs.~\ref{sec:theoBack} and~\ref{sec:propMethod}. They are primarily related to the fact that early reflections in an SRIR are both impulsive, with very limited temporal support, and highly correlated with the direct sound, and consequently with each other. In contrast, many state-of-the-art localization techniques rely on the estimation of a spatial covariance matrix over extended time windows and assume statistical independence between sources. Furthermore, several approaches do not explicitly address the estimation of the number of active sources, which is a critical requirement for blind analysis of SRIRs.

\par Two widely used parametric approaches for SRIR analysis and reproduction are spatial impulse response rendering (SIRR)~\cite{merimaa_spatial_2005} and the spatial decomposition method (SDM)~\cite{tervo_spatial_2013}. Both techniques typically estimate at most one dominant DoA per analysis frame, although higher-order SIRR (HO-SIRR)~\cite{mccormack_higher-order_2020} extends this framework to allow multiple simultaneous DoAs. SIRR and HO-SIRR rely on the estimation of a pseudo-intensity vector (PIV), which in term requires relatively long analysis windows. Similarily, McCormack et al.\ \cite{mccormack_sharpening_2019} proposed an ``angular sharpening'' approach for steered response power (SRP) localization maps, which also depends on spherical sector-based PIV estimation, in a manner analogous to HO-SIRR.

\par Subspace methods such as multiple signal classification (MUSIC)~\cite{schmidt_multiple_1986} and the estimation of signal parameters via rotational invariance techniques (ESPRIT)~\cite{roy_esprit_1989} enable the localization of multiple simultaneous sources. However, these methods also require covariance matrix estimation, which again presupposes sufficiently long analysis windows. In addition, \mbox{MUSIC}- and \mbox{ESPRIT}-based methods generally assume that the sources are either uncorrelated or, at most, weakly correlated.

\par Nevertheless, several adaptations of \mbox{MUSIC}-inspired methods have been proposed for SRIR analysis. Khaykin and Rafaely~\cite{khaykin_acoustic_2012} introduced a frequency-smoothing approach to estimate the covariance matrix from a single-snapshot full-length SRIR. However, due to the frequency-dependent characteristics of the spherical array response in SMA measurements (see Sec.~\ref{sec:SMAPWD}), the full frequency bandwidth cannot be fully exploited. Furthermore, the maximum number of detectable reflections remains fundamentally limited by the maximum spherical harmonic (SH) order of the SMA in this formulation.

\par To address these limitations, Huleihel and Rafaely~\cite{huleihel_spherical_2013} proposed a windowed time-smoothing strategy to construct a \mbox{MUSIC} pseudospectrum. Morgenstern and Rafaely~\cite{morgenstern_modal_2018} later introduced a modal smoothing technique in the SH (or wave spectral) domain requiring a spherical loudspeaker array (SLA) and therefore restricting its applicability to a multiple-input multiple-output (MIMO) setting \cite{morgenstern_mimo_2017}. Notably, neither approach provides a complete framework for estimating both the number of reflections and their associated DoAs.

\par More recently, Shlomo and Rafaely proposed a blind localization method for early room reflections based on phase-aligned spatial correlation~\cite{shlomo_blind_2021}. Their approach demonstrates the feasibility of estimating multiple reflection directions from SMA-measured SRIRs without requiring prior knowledge of the reflection structure. However, it relies on phase alignment and spatial correlation of the measured signals, whereas the present work formulates the localization problem directly in the SH domain through a Herglotz wavefunction representation.

\subsection{\label{sec:contribution}Overview of the Proposed Approach}

\par Motivated by the limitations of existing approaches, this paper introduces a localization framework based on the Herglotz wavefunction formalism for detecting and characterizing multiple early reflections within a single SRIR analysis frame. The Herglotz representation describes the incident sound field as a continuous angular distribution of plane waves, allowing the localization problem to be formulated as the estimation of a directional function on the sphere rather than as the direct identification of a fixed number of discrete sources.

\par The proposed framework consists of three main steps. First, the measured SMA sound field is transformed into a Herglotz-based localization function in the spherical harmonic domain. Second, peaks in the resulting localization function are detected using a sparse set of spherical Gaussian components, whose parameters provide estimates of the directions and relative strengths of incident reflections. Finally, an adaptive information-criterion-based procedure estimates the number of Gaussian components and, consequently, the number of reflections present in the analysis frame. In contrast to SRP-based analysis, which evaluates the sphere by considering one direction independently at a time, and can therefore suffer from smearing and spurious peaks when highly correlated echoes overlap, the Herglotz approach formulates localization as a global inverse
problem. We therefore solve the global inverse problem for the optimal continuous directional distribution over the entire sphere simultaneously, enabling more accurate resolution of multiple concurrent reflections.

\par Although the long-term objective is to construct complete early-reflection cartographies through frame-by-frame analysis of SRIRs, the present work focuses on the more fundamental problem of localizing and characterizing multiple reflections occurring within a single analysis frame.

\par The paper is structured as follows. Section~\ref{sec:theoBack} introduces the theoretical background underlying the proposed method. After a description of SRIR temporal segmentation in Section~\ref{sec:SRIRtimeSeg}, the sound field generated by the scattering of a plane wave by a rigid sphere is reviewed. Section~\ref{sec:herglotz} introduces the Herglotz representation of sound fields. These formulations are combined in Section~\ref{sec:propMethod}, where the proposed localization framework and echo detection procedure are developed.

\par In Section~\ref{sec:herglotz}, the field generated by multiple independent plane waves is expressed in the spherical harmonic (SH) domain. Section~\ref{sec:echoLocFunc} extends this formulation to the more general case of echo-driven sound fields, where reflections are modeled as a set of localized contributions that may be represented using radial Gaussian basis functions on the sphere. Importantly, the number of echoes is not assumed to be known a priori. Sections~\ref{sec:echoDetect} and \ref{sec:echoNumEst} present the resulting detection procedures and the associated adaptive estimation of the number of echoes. Finally, Section~\ref{sec:validation} presents an evaluation of the proposed method.

\section{\label{sec:theoBack}Theoretical Background}

\subsection{\label{sec:SRIRtimeSeg}SRIR Time Segmentation}

\par Throughout this work, the SRIR is assumed to be temporally segmented into an \emph{early} region, consisting predominantly of the direct sound and its first discrete reflections, followed by a stochastic, exponentially decaying late reverberation tail, as originally described by Sabine, Eyring, and later formalized by Schroeder~\cite{schroeder_new_1965}. The transition between these two regimes is commonly referred to as the \emph{mixing time}, $\tMix$, following the interpretation introduced by Polack~\cite{polack_playing_1993}.

\par The mixing time considered here is defined from the temporal evolution of a measure of the SRIR's spatial incoherence, following the approaches of Götz et al.~\cite{gotz_mixing_2015} and Massé et al.~\cite{masse_robust_2020,masse_denoising_2020}. It is measured relative to ToA of the direct sound, $t_0$, and is therefore specific to each SRIR.

\par A complete analysis of the early reflections consequently requires frame-by-frame characterization over the interval $t\in[t_0,\tMix]$. Such an analysis involves several additional processing steps that are beyond the scope of the present work, including the selection of analysis frames, the rejection of false detections, and the association of repeated detections corresponding to the same physical reflection. Some of these aspects have previously been discussed by Massé~\cite{masse_analysis_2022}.

\subsection{\label{sec:SMAPWD}Plane-Wave Field Measurement with an SMA}
\subsubsection{SH representation of a single plane wave}

\par The total sound field at an arbitrary point $\br = (r,\vOmega) = (r,\theta,\phi)$ outside a \emph{rigid} sphere with radius $r_s$ resulting from a single unit amplitude plane wave incident from direction $\bd = (1,\hat{\vOmega}) = (1, \theta_d,\phi_d)$ is obtained as the sum of incident and scattered contributions:
\begin{equation}
    \label{eq:totPWfield}
    \begin{split}
        u(kr,\br,\bd)&=u_\mathrm{in}(kr,\br,\bd)+u_\mathrm{sc}(kr,\br,\bd)\\
        &=\sum_{\ell=0}^\infty{b_\ell(kr)}\sum_{m=-\ell}^\ell Y_{\ell m}(\vOmega)Y_{\ell m}^*(\hat{\vOmega}),
    \end{split}
\end{equation}
where 
\begin{equation}
  b_\ell(kr)=4\pi{\I^\ell}\left[j_\ell(kr)-\frac{j_\ell'(kr_s)}{h_\ell'(kr_s)}h_\ell(kr)\right]
\end{equation} 
is the so-called mode strength (or holographic function) for a rigid sphere \cite[Eq.~(6.186)]{Williams:1999wm}. The $j_\ell$ and $h_\ell$ denote the spherical Bessel and Hankel functions of order $\ell$, respectively, and the prime denotes differentiation of the corresponding function with respect to its argument.

\par On the other hand, for a function $u$ defined on the unit sphere, the coefficients $u_{\ell m}$ can be obtained from the orthonormality of the spherical harmonics:
$$
u_{\ell m} = \int_{\sphere} u(\vOmega) Y^*_{\ell m}(\vOmega) d\vOmega.
$$
In practice, the sound field $u$ is sampled by the $Q$ transducers of the SMA located at positions $\mathbf{r}_q$ on the sphere with radius $r = r_s$. The theory of optimal sampling on the sphere is beyond the scope of this work; relevant references include Meyer and Elko~\cite{meyer_highly_2002}, Abhayapala~\cite{abhayapala_theory_2002}, Rafaely~\cite{rafaely_analysis_2005,rafaely_spatial_2008}, and Chardon et al.~\cite{Chardon:2015hd}. 

\par For practical applications, the sum over $\ell$ in Eq.~(\ref{eq:totPWfield}) must be truncated at some maximum order $L$. The maximum attainable SH order is determined by factors such as the accuracy of the quadrature scheme and the spatial sampling density; see, e.g., \cite{Chardon:2015hd}. For example, the 32-capsule Eigenmike (EM32) supports accurate SH representations up to $L=4$, whereas the 64-capsule Eigenmike (EM64) is practically limited to approximately $L \approx 6$ due to the poorer conditioning of its encoding matrix. As a simple rule of thumb for determining the required SH order, we refer to \cite[Eq.~(17)]{WarAbh01}.

\subsubsection{Compensation of sphere radius}

\par Importantly, only the mode strength $b_\ell$ depends on the sphere radius. Consequently, if the SH coefficients are compensated by an inverse model of $b_\ell$, the resulting representation becomes independent of the specific SMA geometry \cite{daniel_representation_2001}. In practice, this compensation is achieved by applying encoding (or correction) filters~\cite{meyer_highly_2002,daniel_further_2004} after the SH transform. Denoting the frequency response of these filters by $a_\ell(f)$, the effective modal response becomes $c_\ell(f)=a_\ell(f)b_\ell(f)$, yielding the corrected SH coefficients $\widetilde{u}_{\ell m}(f,\hat{\vOmega})$. The filters are typically designed such that $c_\ell(f)$ is as flat and as geometry-independent as possible over the usable frequency range.

\par In practice, the frequency response $c_\ell(f)$ is not perfectly flat, resulting in a frequency-dependent function (see, for example, Daniel and Moreau~\cite{daniel_further_2004}, among many others):
\begin{equation}
  \label{eq:dirFunc}
  w(f,\vOmega)=\sum_{\ell=0}^L\sum_{m=-\ell}^\ell{c}_\ell(f)Y_{\ell m}(\vOmega)Y_{\ell  m}^*(\hat{\vOmega}).
\end{equation}

\par
\begin{figure}[!h]%
    \centering%
    \includegraphics[width=\linewidth]{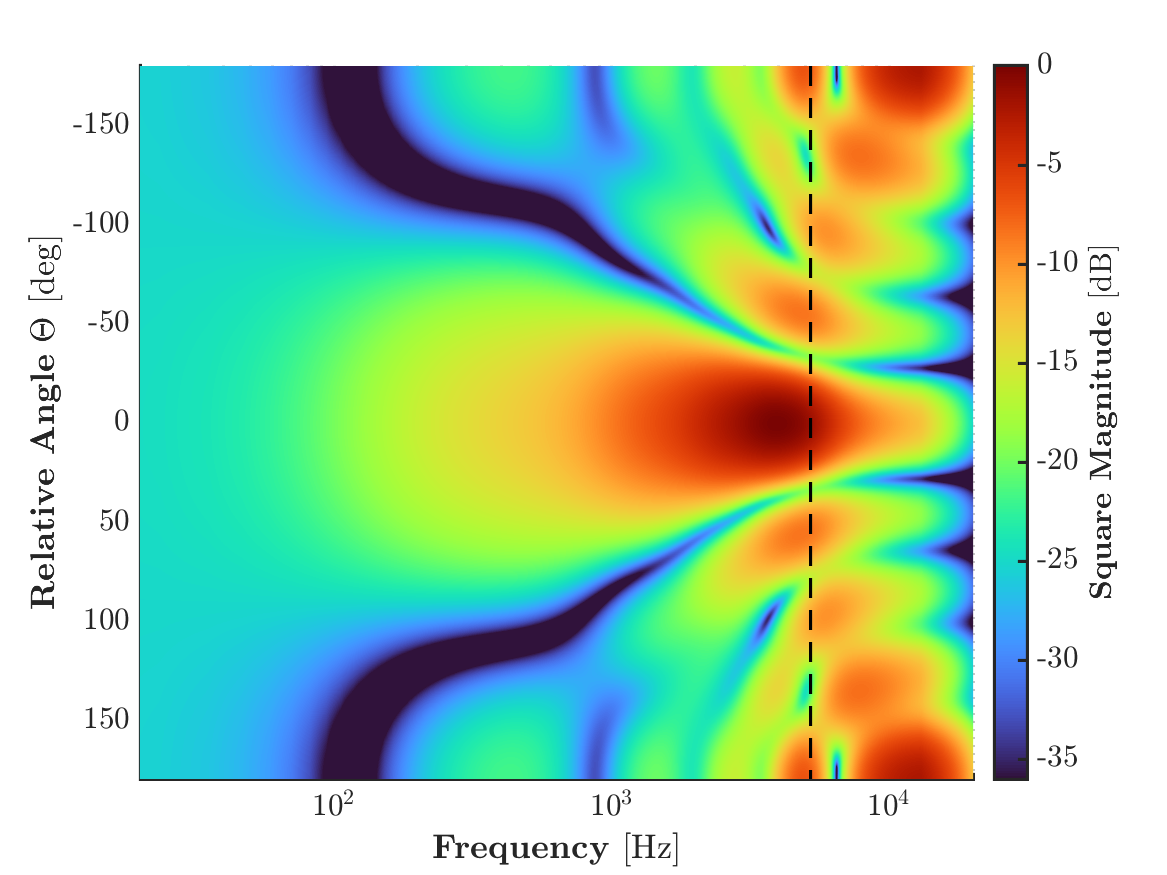}%
    \caption{$w(f,\Theta)$ (Eq.~\ref{eq:dirFunc}) evaluated along the axisymmetric great circle $\Theta$ with respect to the incident direction $\hat{\vOmega}$, for a simulated $L=4$ Higher-Order Ambisonics (HOA) encoding \cite{daniel_further_2004} of a plane wave measured with the mh-acoustics EM32 spherical microphone array (SMA). The dashed line indicates the spatial aliasing frequency, $f_\mathrm{alias}$.}%
    \label{fig:SMAHOAPWDfreqResp}%
\end{figure}

\begin{figure}[!h]
  \centering
  \includegraphics[width=\linewidth]{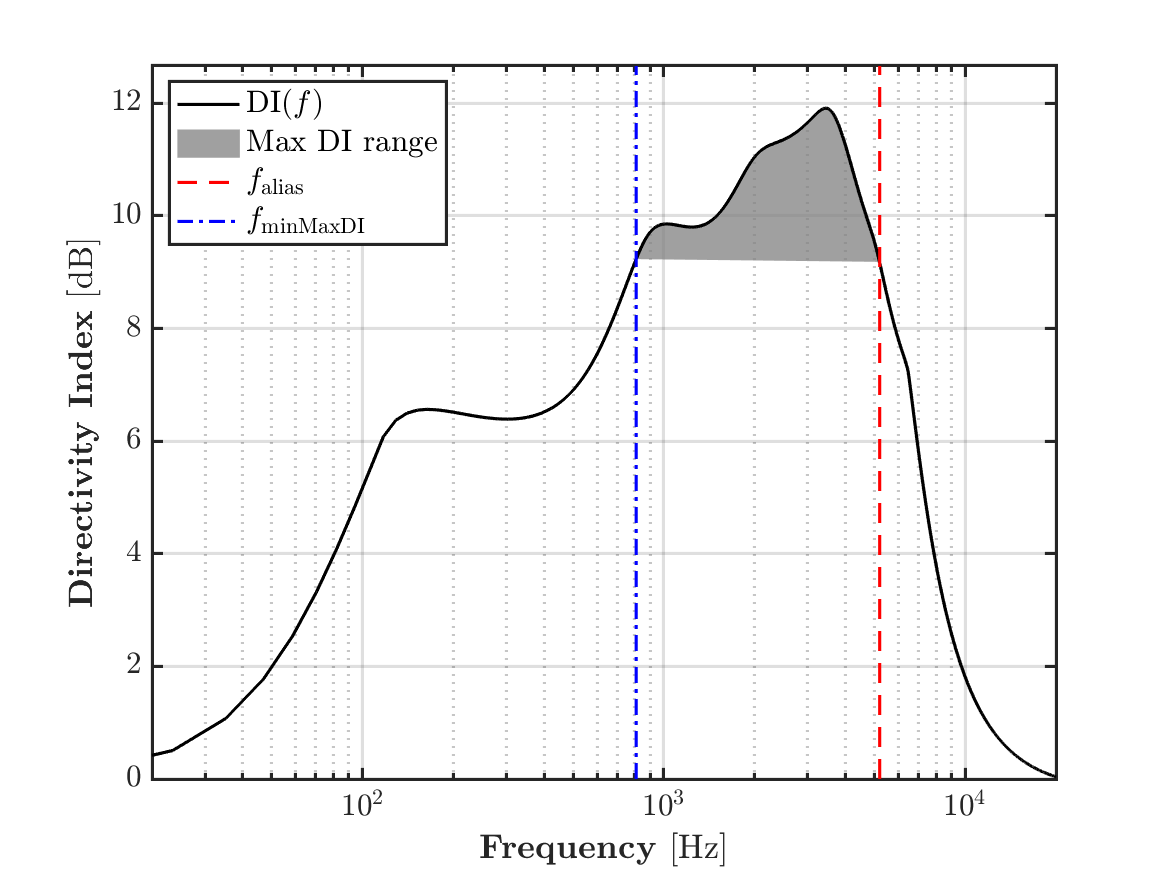}
  \caption{Directivity index (DI) as a function of frequency for the simulated order $L=4$ HOA encoding of a plane wave as measured by the mh-acoustics EM32 SMA. The red dashed line indicates the spatial aliasing frequency $f_\mathrm{alias}$, the blue dash-dotted line the lower bound $f_\mathrm{minMaxDI}$ of the SMA's operational range. The shaded region denots the frequency band of maximal directivity, such that $\mathrm{DI}(f_b)\geq\mathrm{DI}(f_\mathrm{alias})\ \forall\ f_b\in\left[f_\mathrm{minMaxDI},f_\mathrm{alias}\right]$.}
  \label{fig:SMAHOAPWDfreqDI_maxDIrange}
\end{figure}

 Fig.~\ref{fig:SMAHOAPWDfreqResp} shows $w(f,\Theta)$ for an $L=4$ HOA-encoded plane wave, evaluated along the axisymmetric great circle defined with respect to the reference direction $\hat{\vOmega}$. The directivity pattern is simulated for an mh-acoustics EM32, and the corresponding directivity index (DI)~\cite{li_flexible_2007} is shown in Fig.~\ref{fig:SMAHOAPWDfreqDI_maxDIrange}. Based on the DI, a ``maximally directive'' frequency band can be identified. The upper bound is given by the spatial aliasing frequency $f_\mathrm{alias}$~\cite{rafaely_plane-wave_2004}, beyond which the directional response becomes increasingly ambiguous. A lower bound $f_\mathrm{minMaxDI}$ is then defined such that the DI remains at least as high as its value at the aliasing limit, i.e., $\mathrm{DI}\left(f_b\right)\geq\mathrm{DI}\left(f_\mathrm{alias}\right)$ for all $f \in [f_\mathrm{minMaxDI}, f_\mathrm{alias}]$.

\section{\label{sec:propMethod}Proposed Method}
\subsection{Herglotz Wavefunction}\label{sec:herglotz}

\par As discussed in Section~\ref{sec:SRIRtimeSeg}, the predominant early reflections in an SRIR are assumed to be well approximated by ideal impulsive arrivals. Consequently, within a single analysis frame, the sound field measured by an SMA can be interpreted as a superposition of multiple plane waves arriving from different directions, each contributing over the analyzed frequency range. This naturally motivates the use of the Herglotz wavefunction formalism~\cite[p.~60]{colton_inverse_2013}.
The Herglotz wavefunction is defined as
\begin{equation}
\label{eq:contHergWF}
    u(k,\br) = \int\limits_{\sphere} g(\mathbf{d}) e^{ik \br\cdot \mathbf{d}} d\sphere
\end{equation}
where the sound field $u$ at position $\br$ is represented as a continuous superposition of plane waves arriving from all directions $\bd\in\mathbb{S}^2$. The direction-dependent weighting function $g(\bd)$, referred to as the Herglotz kernel, describes the contribution of each plane-wave component as a continuous function. The Herglotz kernel is assumed to satisfy $g\in L^2(\sphere)$, i.e., $g(\bd)$ must be square integrable over the unit sphere.

\par In \cite[Lemma, 3.20]{colton_inverse_2013}, it is shown that if an incident sound field is a Herglotz wavefunction with kernel $g(\bd)$, then the field scattered by a sphere, and hence the resulting total field, can also be expressed as a Herglotz wavefunction with the same kernel 
\begin{equation}\label{eq:totPWfieldII}
u(k,\br) = \int_{\sphere} g(\bd) u^{\text{pw}}(\br,\bd)d\sphere(d),
\end{equation}
where $u^{\text{pw}}(\br,\bd)$ denotes the acoustic field at $\br$ resulting from the scattering of a single plane wave arriving from direction $\bd$.
 
\par This formulation also holds for the sound field evaluated on the sphere. Thus, given measurements of the sound field over the SMA surface, the problem of finding directions of the incident reflections can be formulated as an inverse problem in which the Herglotz kernel is estimated from the measurements. The dominant DoAs can then be identified from the principal maxima of the Herglotz kernel $g$.

Applying the spherical harmonic transform (SHT) to both sides of Eq.~(\ref{eq:totPWfieldII}) yields 
\begin{equation}\label{eq:totPWfieldIISH}
  \begin{split}
    u_{\ell m}(f)=\int\limits_{\sphere} g(\hat{\vOmega}) u^{\text{pw}}_{\ell m}(kr_s,\hat{\vOmega}) d\hat{\vOmega}.
  \end{split}
\end{equation}
Here, the incident direction $\bd$ has been replaced by $\hat{\vOmega}$ to emphasize that the integration is performed over all possible plane-wave directions.

\par The estimation of the Herglotz kernel $g(\hat{\vOmega})$ requires a numerical discretization of the integral over the unit sphere. This differs subtly from conventional beamforming methods, which typically evaluate the sound filed over a fixed grid of scanning directions. In the Herglotz formulation, the discretization of the unit sphere is determined solely by the accuracy required for numerical quadrature. 

\par Furthermore, since the proposed method operates on HOA-encoded SRIRs, the SH coefficients are assumed to have been compensated by the encoding filters described in Section~\ref{sec:SMAPWD}. Consequently, the mode strengths $b_\ell(f)$ are replaced by the effective frequency responses $c_\ell(f)$. This compensation is indicated in the following by using $\tilde{u}_{\ell m}$ instead of $u_{\ell m}$ to denote the spherical harmonics coefficients.

\par The sampled kernel provides a directional interpolation of the incident field, where a single echo corresponds to a localized contribution in this domain.
The relationship between the measured SH coefficients and the sampled kernel can be written in matrix form as
\begin{equation}\label{eq:simplecase}
    \tilde{\mathbf{u}}(f)=\mathbf{D}(f)\mathbf{g},
\end{equation}
where the $(n,j)$-th entry of the dictionary matrix $\mathbf{D}(f)$ is given by 
\[
\mathbf{D}_{n j}=\beta_j c_\ell(f)Y_{\ell m}^*(\hat{\vOmega}_j).
\] 
The multi-index 
\[
n=\ell^2+\ell+m+1,\ \ell = 0,\dots,L,\; m = -\ell, \dots, \ell
\] 
enumerates all spherical harmonics up to order $L$ (cf. ~\cite{abhayapala_theory_2002,noisternig_reconstructing_2011}). The quantities
$\beta_j$ and $\hat{\vOmega}_j$ denote the $D$ quadrature weights and nodes associated with the discretization of the integral over the sphere in Eqs.~(\ref{eq:totPWfieldII}) and (\ref{eq:totPWfieldIISH}), respectively. The vector
\[
\tilde{\bu}(f) = [\tilde{u}_{00}(f),\dots,\tilde{u}_{LL}(f)]^\top
\]
contains the modified spherical harmonic coefficients, while 
\[
\mathbf{g}=[g(\hat{\vOmega}_1),\dots,g(\hat{\vOmega}_D)]^\top.
\]
contains the sampled values of the Herglotz kernel.

\subsection{\label{sec:echoLocFunc}Echo Localization Function}
\par If the early reflections were simply modeled as a superposition of independent plane waves, it would be sufficient to determine $\mathbf{g}$ based on Eq.~(\ref{eq:simplecase}). However, the early segment of an SRIR generally consists of multiple reflections that overlap within a given analysis frame and cannot necessarily be interpreted as independent plane-wave components. We therefore assume that each analysis frame in the early segment, $t\in[t_0,t_\mathrm{mix}]$, contains $P$ impulsive echoes, where each echo can be represented by a Herglotz wavefunction with an associated kernel that is invariant with respect to frequency.
Each echo is characterized by a frequency-dependent amplitude $a_p(f)$ and a discretized Herglotz kernel 
\[
\mathbf{g}_p = [g_p(\hat{\vOmega}_1),\dots,g_p(\hat{\vOmega}_D)]^\top, \quad p = 1,\dots,P.
\]
The resulting SH coefficients of the encoded sound field are therefore given by
\begin{equation}
  \label{eq:multiHerg}
  \tilde{u}_{\ell m}(f)=c_\ell(f)\sum_{j=1}^D\beta_jY_{\ell m}^*(\hat{\vOmega}_j)\sum_{p=1}^P a_p(f)g_p(\hat{\vOmega}_j).
\end{equation}

\par Equivalently, this expression can be written in matrix form as
\begin{equation}
  \label{eq:hergLocMat}
  \begin{split}
    \tilde{\mathbf{u}}(f)&=\mathbf{D}(f)\mathbf{G}\mathbf{a}(f)\\
    &=\mathbf{D}(f)\mathbf{m}(f),
  \end{split}
\end{equation}
where $\mathbf{G}_{ij}=\mathbf{g}_j(\hat{\vOmega}_i)$ contains the discretized kernels of the individual echoes, $\mathbf{a}(f)$ contains the frequency-dependent amplitudes $a_p(f)$, and 
$
\mathbf{m}(f)=\mathbf{G}\mathbf{a}(f)
$ 
is a vector of length $D$ that will henceforth be referred to as the \textit{localization function}. This function combines the directional contributions of all echoes and therefore contains the information required to identify their DOAs. Importantly, the number of echoes $P$ is only implicitly "encoded" in $\mathbf{m}$ and need not to be known a priori; instead, it is estimated adaptively as described in Section~\ref{sec:echoNumEst}.

\par The first step of the proposed method is to estimate the localization function $\mathbf{m}(f)$ from the SH coefficients measured by the SMA. 
The dictionary matrix $\mathbf{D}(f)$ has $(L+1)^2$ rows and $D$ columns, where $D$ denotes the number of quadrature nodes on the sphere  used to interpolate the Herglotz kernel.

\par We estimate the localization function using Tikhonov regularization~\cite{hansen_regularization_1998}:
\begin{equation}
    \label{eq:hergLocEst}
    {\mathbf{m}}(f)=\left(\mathbf{D}^\mathrm{H}\mathbf{D}+\mathbf{T}^\top\mathbf{T}\right)^{-1}\mathbf{D}^\mathrm{H}\tilde{\mathbf{u}}(f),
\end{equation}
where $\mathbf{T}=\mu\mathbf{I}^{D\times{D}}$ is the regularization operator, $\mu$ is the regularization parameter, and $\mathbf{I}^{D\times{D}}$ the $D\times{D}$ identity matrix. This formulation favors solutions with reduced norm and stabilizes the inversion of the ill-conditioned system.

\par In this formulation, the regularization operator effectively acts as a spatial smoothing constraint on the estimated localization function. This is consistent with the smoothness properties of the Herglotz kernel and implies that, although a finer sampling scheme may be used to represent the solution, the achievable spatial resolution remains fundamentally limited by the maximum SH order.

\par Finally, the frequency dependence has so far been treated continuously for notational simplicity. In practice, frame-based analysis (e.g., using a short-time Fourier transform, STFT) discretizes the frequency axis into $K$ bins $f_k$ over the positive-frequency range. The localization function could therefore be estimated independently for each frequency bin using Eq.~(\ref{eq:hergLocEst}). However, as discussed in Section~\ref{sec:SMAPWD}, the directivity of an HOA-encoded plane wave depends strongly on frequency, and directional information becomes unreliable above the spatial aliasing frequency $f_\mathrm{alias}$. Consequently, estimation of $\mathbf{m}(f_k)$ is restricted to the maximally directive frequency band defined in Section~\ref{sec:SMAPWD} and illustrated in Fig.~\ref{fig:SMAHOAPWDfreqDI_maxDIrange}, i.e.\ for $f_k\in\left[f_\mathrm{minMaxDI},f_\mathrm{alias}\right]$ such that $\mathrm{DI}(f_k)\geq\mathrm{DI}(f_\mathrm{alias})$.

\subsection{\label{sec:echoDetect}Echo Detection}

\par The next step consists of identifying the directional maxima of the localization function ${\mathbf{m}}(f_k)$ in order to estimate the DOAs of the incident echoes present within the current analysis frame. Since the magnitude of each localized contribution is related to the corresponding echo strength, this procedure also provides a first characterization of their relative amplitudes.

\par In general, the localization function ${\mathbf{m}}(f_k)$ is complex-valued, since the HOA-encoded response $c_\ell(f)$, the SH basis functions, and the STFT coefficients $\tilde{\mathbf{u}}(f)$ are complex valued quantities. However, the directional information indicating the presence of an echo is contained in the magnitude of the localization function rather than its phase. Furthermore, recall that the Herglotz kernels $g_p(\hat{\vOmega})$ associated with the individual echoes are assumed to be frequency invariant. Consequently, the detection procedure should combine information across frequency while remaining independent of the particular frequency bin. We define a frequency-averaged localization function as

\begin{equation}
  \label{eq:hergLocFunc}
  \overline{m}_d=\frac{1}{\Gamma_\mathrm{tot}}\sum_k\gamma_k\left|{m}_{d}(f_k)\right|,
\end{equation}
where the summation is performed over the maximal directive frequency range $f_k\in\left[f_\mathrm{minMaxDI},f_\mathrm{alias}\right]$, and where 
\[
\gamma_k=\max_d\left(\left|{m}_{d}(f_k)\right|\right), \quad \Gamma_\mathrm{tot}=\sum_k\gamma_k.
\] 
The weighting factor $\gamma_k$ emphasizes frequency bins with stronger directional information, while reducing the influence of bins where the localization function is dominated by noise or insufficient directivity. The index $d$ denotes the $d$-th sampled direction of the localization grid.

\par Having defined the frequency-independent localization function $\overline{\mathbf{m}}$, we now formulate the detection of its directional peaks as a parametric estimation problem. We model the localization function as a sum of $P$ spherical radial basis functions (RBFs), using spherical Gaussian functions to represent the contribution of each echo:
\begin{equation}
    \label{eq:locFuncModel}
    \widehat{m}_d\left(\mathbf{\Psi}\right)=\sum_p\sqrt{h_p^2}e^{\lambda_p\left(\mathbf{d}_p\cdot\mathbf{r}_d-1\right)},
\end{equation}
where $\mathbf{d}_p$ denotes the position of the $p$-th Gaussian peak on the unit sphere, $h_p$ its amplitude coefficient (relative presence), and $\lambda_p$ its angular concentration parameter (lobe sharpness). The vector $\mathbf{r}_d=\left[1,\vOmega_d\right]$ denotes the direction associated with the $d$-th sample of the localization grid.

The model parameters are collected in the matrix $\mathbf{\Psi}\in\mathbb{R}^{4\times P}$, whose $p$-th column is defined as
\[
\boldsymbol{\psi}_p
=
\left[\vOmega_p,h_p,\lambda_p\right]^\top.
\]
The estimation problem therefore consists of finding the parameters $\mathbf{\Psi}$ that best describe the measured localization function. In particular, the estimated directions $\vOmega_p$ provide the DoAs of the detected echoes, while the coefficients $h_p$ provide a relative measure of their strengths.

\par The model can equivalently be written in matrix form as
\begin{equation}
    \label{eq:matLocFuncModel}
    \widehat{\mathbf{m}}\left(\mathbf{\Psi}\right)=\mathbf{M}\mathbf{h},
\end{equation}
where $\mathbf{h}$ contains the presence factors $\sqrt{h_p^2}$ and the entries of $M_{d,p}$ are given by $M_{d,p}=e^{\lambda_p\left(\mathbf{d}_p\cdot\mathbf{r}_d-1\right)}$.

The use of $\sqrt{h_p^2}$ ensures that the amplitude coefficients remain nonnegative during the optimization.

\par The parameters $\vOmega_p$, $h_p$, and $\lambda_p$ are estimated by minimizing the weighted least-squares objective
\begin{equation}
  \label{eq:gradDescCost}
  \mathcal{L}\left(\mathbf{\Psi}\right)=\frac{1}{\mathcal{M}_\mathrm{tot}}\sum_d\overline{m}_d\left[\overline{m}_d-\widehat{m}_d\left(\mathbf{\Psi}\right)\right]^2,
\end{equation}
where $\mathcal{M}_\mathrm{tot}=\sum_d\overline{m}_d$. 
This corresponds to a mean squared error (through a gradient descent on the linear system of equations described by Eq.~(\ref{eq:matLocFuncModel})) weighted by the measured localization function $\overline{\mathbf{m}}$ itself. Consequently, errors in regions containing strong localization peaks contribute more strongly to the objective than errors in regions where no significant directional energy is present. This weighting is desirable because the objective is primarily concerned with accurately fitting the detected peaks, while avoiding over-fitting to background fluctuations in low-energy regions.

\par The optimization is performed using gradient descent with an adaptive step size determined through a secant-based estimate following the Barzilai--Borwein method~\cite{barzilai_two-point_1988}.

\subsection{\label{sec:echoNumEst}Estimation of the Number of Incident Echoes}

\par The localization procedure described above assumes that the number of peaks $P$ is known when modelling the localization function. In practice, however, this quantity is unknown when analysing an SRIR frame in a blind manner. An additional procedure is therefore required to determine the appropriate number of Gaussian components to include in the model.

\par We propose an iterative estimation strategy based on an information criterion inspired by the Akaike information criterion (AIC)~\cite{akaike_new_1974}:
\begin{equation}
  \label{eq:AIC}
  \mathcal{C}=-\left\{\ln\left[\mathcal{L}\left(\mathbf{\Psi}\right)\right]+P\right\},
\end{equation}
where $\ln(\cdot)$ denotes the natural logarithm. This criterion introduces a trade-off between the quality of the fit (absolute fitting error) and the complexity of the model. Increasing the number of Gaussian components generally reduces the reconstruction error, but additional components may simply model spatial fluctuations or noise rather than physically (and also perceptually) meaningful reflections. The use of an information criterion can, in principle, constrain the modelling process such that the selected number of radial Gaussian functions corresponds to the actual number of incident echoes. In practice, this relies on the assumption that physically meaningful reflections provide a sufficient reduction in fitting error to justify their inclusion, whereas additional components primarily capture spatial fluctuations or noise.

\par The criterion above was selected empirically based on its performance in the preliminary simulations presented in Section~\ref{sec:validation}. A systematic comparison with alternative information criteria, such as the Bayesian information criterion (BIC)~\cite{schwarz_estimating_1978}, is beyond the scope of this work. Nevertheless, the influence of the chosen criterion on the detection performance is briefly examined in Section~\ref{sec:echoDetectEval}.

\par The estimation procedure begins by fitting the localization function using a single Gaussian component, i.e.\ $P_0=1$, and evaluating the corresponding criterion value $\mathcal{C}_0$. The number of components is then increased iteratively according to $P_{i+1}=P_i+1$ with the localization function re-estimated and the corresponding criterion $\mathcal{C}_{i+1}$ evaluated at each iteration. As additional Gaussian components are introduced, the criterion initially decreases as the fit improves, before increasing once the reduction in fitting error no longer justifies the increased model complexity. The iterative procedure is therefore terminated when $\mathcal{C}_{i+1}>\mathcal{C}_i$ and the model obtained at the previous iteration is selected as the final estimate of the number of incident echoes. The performance of this procedure is evaluated in Section~\ref{sec:echoDetectEval}.

\section{Evaluation}\label{sec:validation}

\par This section evaluates the proposed method through both simulated and measured data. The objectives are twofold. First, the proposed Herglotz localization function, defined in Eq.~(\ref{eq:hergLocFunc}), is compared with localization maps obtained using conventional SRP and MUSIC approaches. Second, the complete echo detection framework introduced in Sections~\ref{sec:echoDetect} and~\ref{sec:echoNumEst} is evaluated in terms of its ability to estimate the directions, relative amplitudes, and number of incident echoes.

\par To ensure a fair comparison, the same peak detection procedure is applied to every localization map. Specifically, the radial Gaussian model of Eqs.~(\ref{eq:locFuncModel}) and~(\ref{eq:matLocFuncModel}), the gradient-based parameter estimation of Eq.~(\ref{eq:gradDescCost}), and the iterative AIC-based estimation of the number of peaks are used irrespective of the localization method. Since only a limited number of methods in the literature explicitly address peak detection and/or source number estimation, the comparisons presented below focus primarily on the quality of the underlying localization functions rather than on differences introduced by the peak extraction procedure.

\subsection{\label{sec:testSignals}Simulation Setup}

\par The proposed method is first evaluated using simulated analysis frames representative of the early portion of an SRIR, i.e.\ $t\in[t_0,t_\mathrm{mix}]$. Each frame consists of 256 samples ($5.33$~ms at a sampling frequency of $48$~kHz) and contains a prescribed number $P_\mathrm{sim}$ of impulsive reflections.

\par Unless otherwise stated, four echoes are synthesized with incident directions $\vOmega_1=[-120^\circ,45.0^\circ]$, 
$\vOmega_2=[22.5^\circ,60.0^\circ]$, 
$\vOmega_3=[90.0^\circ,-22.5^\circ]$, and 
$\vOmega_4=[-11.25^\circ,-30^\circ]$.
Their relative amplitudes are $0$, $-6$, $-2.5$, and $-12$~dB with respect to the first echo.

\par Each echo is synthesized as temporal impulse (i.e.\ discrete Dirac distribution) bandpass-filtered between $125$~Hz and $16$~kHz. The lower cutoff avoids temporal aliasing caused by long low-frequency impulse responses within the finite analysis window. This bandwidth intentionally exceeds the maximally directive frequency band of the EM32 ($\approx800$--$5200$~Hz) in order to preserve the broadband characteristics of impulsive reflections.

\par An SMA measurement is simulated according to Eq.~(\ref{eq:totPWfield}) using the geometry of the mh-acoustics Eigenmike. Unless otherwise stated, the simulations presented in this section use the EM32, while additional results obtained with the EM64 are presented later to evaluate the proposed method with a higher-order SMA. The simulated sound field is given by
\begin{equation}
  \begin{split}
      u_\mathrm{sim}(f,\vOmega_q)=&\sum_{\ell=0}^{L_\mathrm{sim}}{b}_\ell(f)\sum_{m=-\ell}^\ell{Y}_{\ell m}(\vOmega_q)\\ &\times\sum_{p=1}^{P_\mathrm{sim}}u_p(f)Y_{\ell m}^*(\vOmega_p'),
  \end{split}
\end{equation}
where $\vOmega_q$ denotes the angular positions of the $q$-th microphone.

\par The simulated microphone signals are subsequently encoded into HOA coefficients using an SH transform %equivalent to Eq.~(\ref{eq:totPWfieldSHtrans})
followed by the application of the encoding filters $a_\ell(f)$ described in Sec.~\ref{sec:SMAPWD}~\cite{daniel_further_2004}. The sound field is simulated with a truncation order of $L_\mathrm{sim}=20$, which is sufficiently high to avoid spatial aliasing below the temporal Nyquist frequency for the simulated Eigenmike geometry ($r_s=4.2$~cm) at a sampling rate of $48$~kHz. Consequently, the spatial aliasing observed in the simulated measurements is solely due to the finite spatial sampling of the array, and in particular to the truncation to $L=4$ imposed by the $Q=32$ microphone capsules of the Eigenmike.

\subsection{Localization Function Estimation}

\par
\begin{figure*}[!h]
  \centering
    \subfigure[Herglotz]{\includegraphics[width=0.5\textwidth]{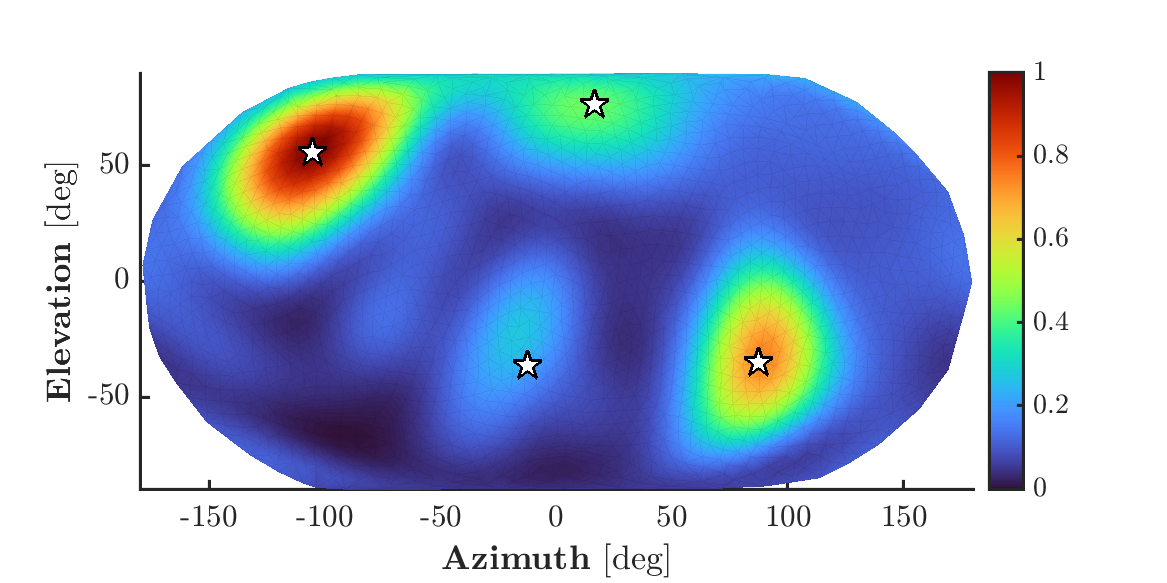}\label{fig:locFunc_herglotz}}%
    \subfigure[SRP]{\includegraphics[width=0.5\textwidth]{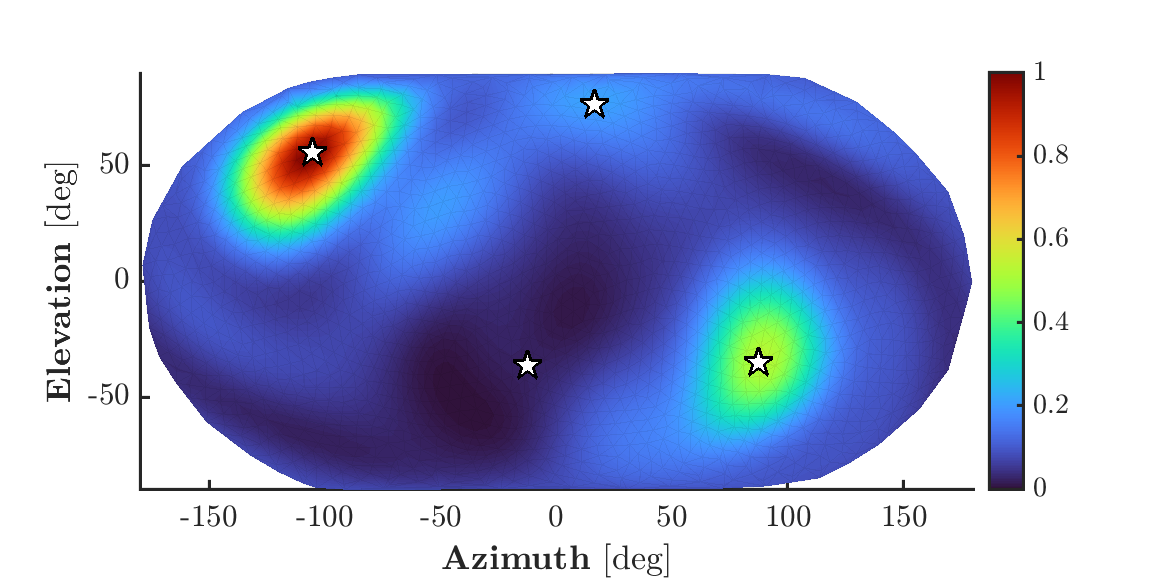}\label{fig:locFunc_origSRP}}\\
    \subfigure[maxWDI]{\includegraphics[width=0.5\textwidth]{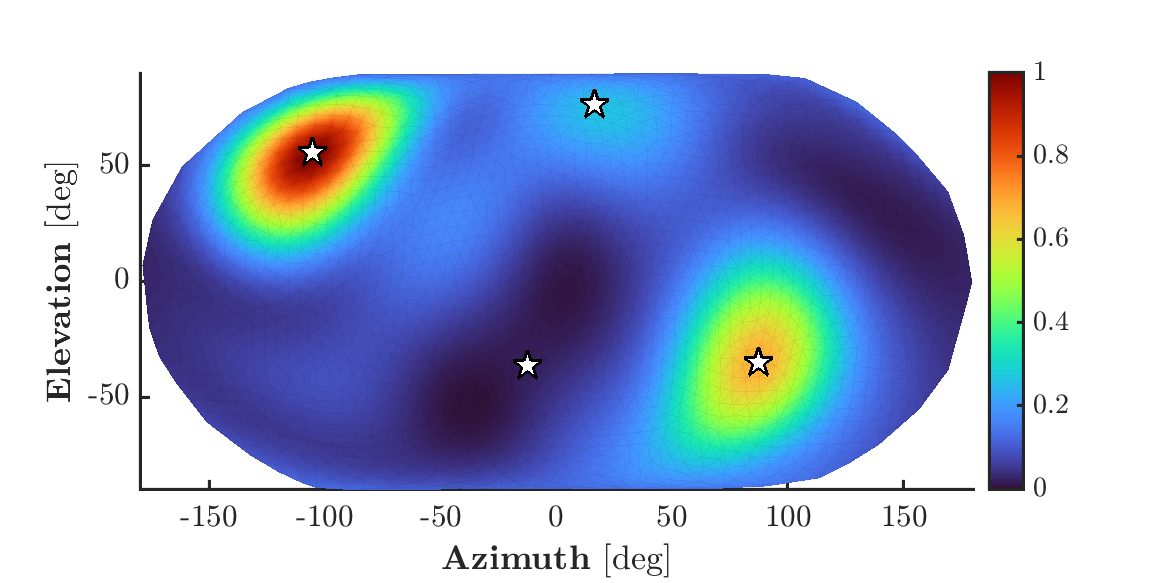}\label{fig:locFunc_SRPmaxWDI}}%
    \subfigure[MUSIC]{\includegraphics[width=0.5\textwidth]{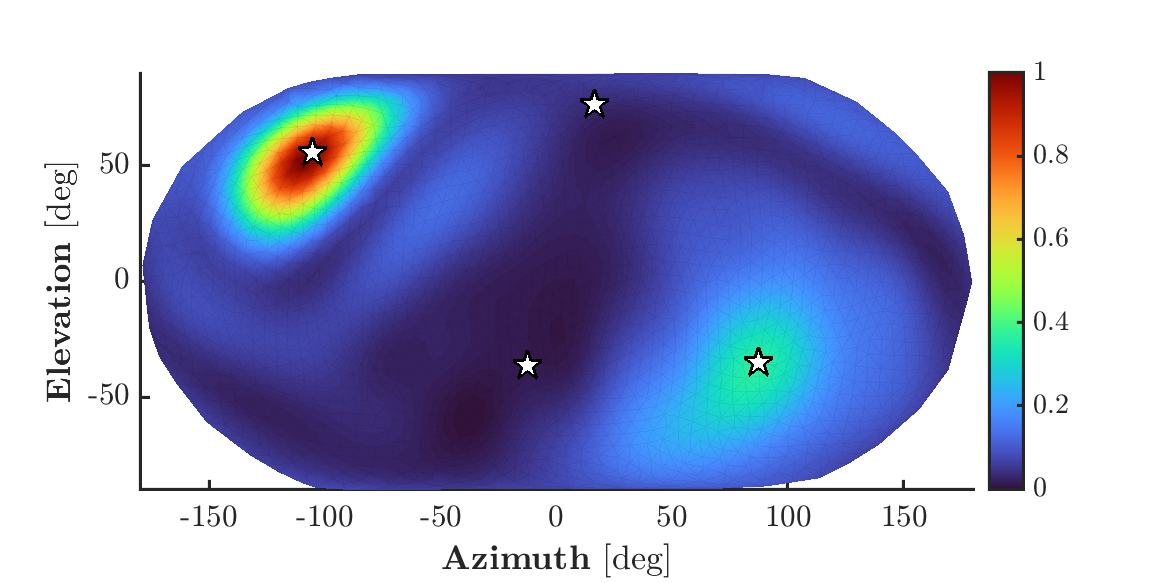}\label{fig:locFunc_MUSIC}}
  \caption{Localization function $\overline{\mathbf{m}}$ calculated according to Eq.~(\ref{eq:hergLocFunc}) for a simulated analysis frame containing $P_\mathrm{sim}=4$ echoes. The echoes are located at DOAs $\vOmega_1=[-120^\circ,45.0^\circ]$, $\vOmega_2=[22.5^\circ,60.0^\circ]$, $\vOmega_3=[90.0^\circ,-22.5^\circ]$, and $\vOmega_4=[-11.25^\circ,-30^\circ]$, respectively. The relative amplitides of the second, third, and fourth echoes are set to $-6$~dB, $-2.5$~dB, and $-12$~dB with respect to the first echo. The spherical data are visualized using the Equal Earth projection~\cite{savric_equal_2019}.}
  \label{fig:locFuncP2}
\end{figure*}

 Fig.~\ref{fig:locFuncP2} compares the final localization functions $\overline{\mathbf{m}}$ obtained from Eq.~(\ref{eq:hergLocFunc}) using four different approaches to estimate the underlying frequency-dependent functions ${\mathbf{m}}(f_k)$ for the test signals described in Sec.~\ref{sec:testSignals}. The proposed Herglotz-based localization function, obtained from Eq.~(\ref{eq:hergLocEst}), is shown in Fig.~\ref{fig:locFunc_herglotz}. For comparison, Fig.~\ref{fig:locFunc_origSRP} shows a conventional SRP localization function based on a plane-wave decomposition (PWD) beamformer~\cite{rafaely_plane-wave_2004}, Fig.~\ref{fig:locFunc_SRPmaxWDI} shows an SRP localization function obtained using a maximum weighted directivity index (maxWDI) beamformer~\cite{masse_analysis_2022}, equivalent to the max-$r_E$ beamformer described in~\cite{zotter_energy-preserving_2012}, and Fig.~\ref{fig:locFunc_MUSIC} shows the \mbox{MUSIC} pseudospectrum~\cite{huleihel_spherical_2013}. The dimension of the \mbox{MUSIC} signal subspace was determined through eigenspectrum analysis following~\cite{morgenstern_modal_2018}, resulting in an estimated number of $\tilde{P}_\mathrm{MUSIC}=2$ sources.

\par The white stars indicate the true simulated echo DoAs. All spherical maps are visualized using the Equal Earth projection introduced by Šavrič et al.~\cite{savric_equal_2019}.

\subsection{\label{sec:echoDetectEval}Echo Detection}

\par
\begin{figure}[!h]
  \centering
  \includegraphics[width=\linewidth]{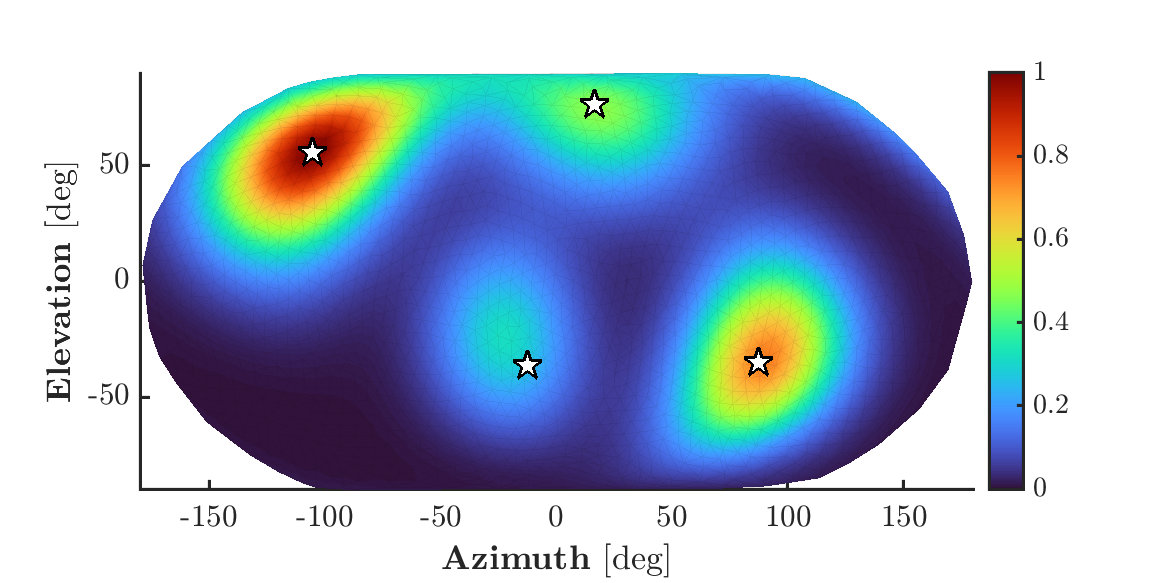}
  \caption{Reconstructed localization function $\widehat{\mathbf{m}}\left(\mathbf{\Psi}\right)$ obtained from the radial Gaussian model of Eqs.~(\ref{eq:locFuncModel}) and~(\ref{eq:matLocFuncModel}) for the simulated four-echo frame described in Sec.~\ref{sec:testSignals}. The $P_\mathrm{sim}=4$ echoes are placed at $\vOmega_1=[-120^\circ,45.0^\circ]$, $\vOmega_2=[22.5^\circ,60.0^\circ]$, $\vOmega_3=[90.0^\circ,-22.5^\circ]$, and $\vOmega_4=[-11.25^\circ,-30^\circ]$, respectively. The model parameters are estimated through gradient descent minimizing the objective function $\mathcal{L}\left(\mathbf{\Psi}\right)$ given in Eq.~(\ref{eq:gradDescCost}). The white stars indicate the true simulated echo DoAs, and the spherical map is visualized using the Equal Earth projection~\cite{savric_equal_2019}.}
  \label{fig:peakDetectP2}
\end{figure}

 Fig.~\ref{fig:peakDetectP2} shows the radial Gaussian model $\widehat{\mathbf{m}}(\mathbf{\Psi})$ obtained from Eqs.~(\ref{eq:locFuncModel}) and~(\ref{eq:matLocFuncModel}) fitted to the proposed Herglotz localization function $\overline{\mathbf{m}}$ [Eq.~(\ref{eq:hergLocFunc}), Fig.~\ref{fig:locFunc_herglotz}] for the simulated test frame described above. The white stars indicate the true simulated echo DoAs, and the spherical map is visualized using the Equal Earth projection~\cite{savric_equal_2019}.

\par Applying the proposed Gaussian fitting and model selection procedure to the Herglotz localization function correctly identifies the $P=4$ simulated echoes in this example. The detection accuracy is evaluated using two complementary metrics. The first is the angular error between the synthesized DoA $\vOmega_p$ and the estimated direction $\widetilde{\vOmega}_p$, obtained from the fitted unit vector $\widetilde{\mathbf{d}}_p$ in Eq.~(\ref{eq:locFuncModel}). The second is the relative presence error, which compares the estimated presence factors $\tilde{h}_p$ with the synthesized relative impulse levels after normalization with respect to the strongest reflection. For this example, the reference levels are $0$, $-6$, $-2.5$, and $-12$~dB. The resulting errors for the Herglotz localization function are reported in the first row of Tab.~\ref{tab:peakDetRes}.

\par Table~\ref{tab:peakDetRes} summarizes the results obtained for the same simulated frame using the three reference localization functions (SRP, maxWDI, and \mbox{MUSIC}) along with the same radial Gaussian fitting and model selection procedure. The final column reports the detection rate for each method, defined as the ratio between the number of detected echoes and the number of synthesized echoes, expressed as a percentage.
\begin{table*}[!h]
  \caption{Peak Localization Accuracy\label{tab:peakDetRes}}
  \centering
  \begin{tabular}{|l||c|c|c|c||c|c|c|c||c|}
    \hline
    \rule{0pt}{2.8ex}\textbf{Method} & \multicolumn{4}{c||}{\textbf{Angular Errors} [deg]} & \multicolumn{4}{c||}{\textbf{Relative Presence Errors} [dB]} & \textbf{Detection Rate} [$\%$]\\
    \hline
    \hline
    \rule{0pt}{2.8ex}Herglotz & $\mathbf{0.773}$ & $2.23$ & $\mathbf{2.93}$ & $\mathbf{15.4}$ & 0 & $\mathbf{0.0919}$ & $\mathbf{0.570}$ & $\mathbf{0.352}$ & $\mathbf{100}$\\
    \hline
    \rule{0pt}{2.8ex}SRP & $0.903$ & $1.25$ & $32.9$ & -- & 0 & $2.27$ & $6.15$ & -- & 75\\
    \hline
    \rule{0pt}{2.8ex}maxWDI & $1.04$ & $\mathbf{0.667}$ & $6.12$ & $53.3$ & 0 & $0.804$ & $2.98$ & $1.95$ & 125\\
    \hline
    \rule{0pt}{2.8ex}\mbox{MUSIC} & $1.35$ & $4.97$ & -- & -- & 0 & $5.60$ & -- & -- & 50\\
    \hline
  \end{tabular}
\end{table*}

\begin{table*}[!h]
  \caption{Peak Localization Accuracy, EM64 Simulation\label{tab:peakDetSimRes}}
  \centering
  \begin{tabular}{|l||c|c|c|c|c|c||c|c|c|c|c|c||c|}
    \hline
    \rule{0pt}{2.8ex} \textbf{Method} & \multicolumn{6}{c||}{\textbf{Angular Errors} [deg]} & \multicolumn{6}{c||}{\textbf{Relative Presence Errors} [dB]} & \textbf{Detection Rate} [$\%$]\\
    \hline
    \hline
    \rule{0pt}{2.8ex} Herglotz & $\mathbf{0.14}$ & $\mathbf{0.729}$ & $\mathbf{0.427}$ & $\mathbf{1.1624}$ & $\mathbf{0.9169}$ & -- & 0 & $\mathbf{0.002}$ & $0.405$ & $0.92$ & $\mathbf{0.426}$ & -- & $\mathbf{83}$\\
    \hline
    \rule{0pt}{2.8ex} SRP & $0.403$ & $1.573$ & $3.024$ & -- & $4.475$ & -- & 0 & $1.609$ & $2.027$ & -- & $1.021$ & -- & 67\\
    \hline
    \rule{0pt}{2.8ex} maxWDI & $3.328$ & $3.821$ & $3.571$ & $10.6826$ & $8.916$ & -- & 0 & $2.143$ & $3.8631$ & $\mathbf{0.319}$ & $3.165$ & -- & 150\\
    \hline
    \rule{0pt}{2.8ex} \mbox{MUSIC} & $1.851$ & $1.622$ & $3.636$ & -- & -- & -- & 0 & $2.81$ & $\mathbf{0.058}$ & -- & -- & -- & 50\\
    \hline
  \end{tabular}
\end{table*}

\par The results indicate that, for the simulated SMA measurement frame containing multiple discrete echoes described in Sec.~\ref{sec:testSignals}, the Herglotz-based localization function provides the most accurate overall reconstruction among the methods considered. It achieves the lowest angular errors for three of the four reflections and consistently yields the smallest relative presence errors. The second reflection, synthesized at $-6$~dB relative to the strongest echo, represents the only case where another method (maxWDI) provides a lower angular error.

\par Furthermore, the proposed localization function is the only one for which the subsequent Gaussian fitting and model selection procedure correctly estimates the number of incident echoes in this example. This result should, however, be interpreted in relation to the information criterion (IC) used in the iterative procedure described in Sec.~\ref{sec:echoNumEst}, since the estimated model complexity depends on the chosen balance between fitting accuracy and parsimony. Inspection of the Herglotz localization map in Fig.~\ref{fig:locFunc_herglotz} also suggests that the detection performance is ultimately limited by the dynamic range between incident echoes. As the level difference increases, weaker reflections may become indistinguishable from spatial noise or spatial aliasing artefacts generated by stronger components. Nevertheless, among the localization functions compared in Fig.~\ref{fig:locFuncP2}, the Herglotz formulation provides the closest representation of the four simulated echo directions.

\subsection{\label{sec:higherOrderSMA}Validation with a Higher-Order SMA}

\par To assess the sensitivity of the proposed method to the spatial order of the measurement array, the simulation described in Sec.~\ref{sec:testSignals} was repeated using the geometry of the mh-acoustics EM64 ($Q=64$ microphone capsules, truncation order $L=6$) rather than the EM32. The number of synthesized echoes was increased to $P_\mathrm{sim}=6$, placed at $\vOmega_1=[-120^\circ,45.0^\circ]$, $\vOmega_2=[22.5^\circ,60.0^\circ]$, $\vOmega_3=[90.0^\circ,-22.5^\circ]$, $\vOmega_4=[-87.0^\circ,-60.0^\circ]$, $\vOmega_5=[-55.0^\circ,12.0^\circ]$, and $\vOmega_6=[0^\circ,0^\circ]$, with relative amplitudes of $0$, $-2.5$, $-4.5$, $-7$, $-10.5$, and $-16.5$~dB with respect to the first echo.

\par
\begin{figure*}[p]
  \centering
    \subfigure[Herglotz]{\includegraphics[width=0.5\textwidth]{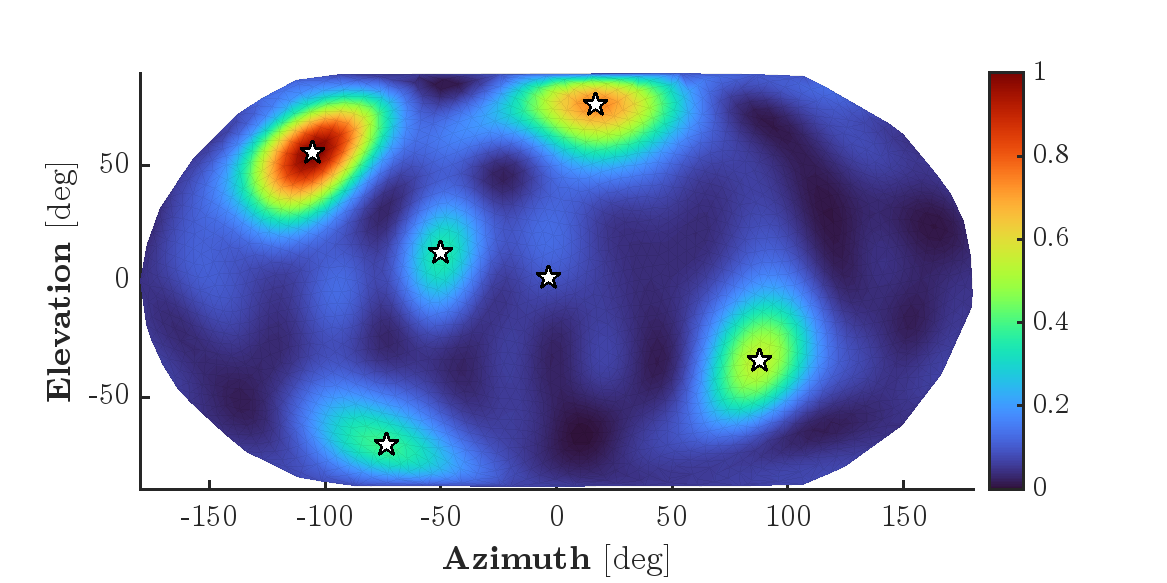}\label{fig:Sim6_herglotz}}%
    \subfigure[SRP]{\includegraphics[width=0.5\textwidth]{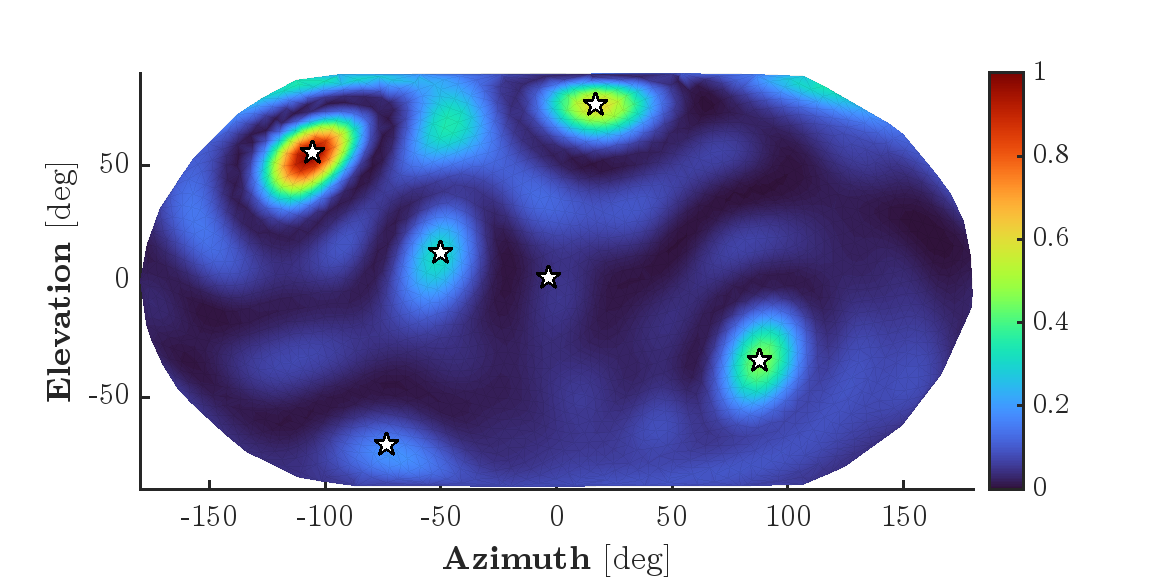}\label{fig:Sim6_SRP}}\\
    \subfigure[maxWDI]{\includegraphics[width=0.5\textwidth]{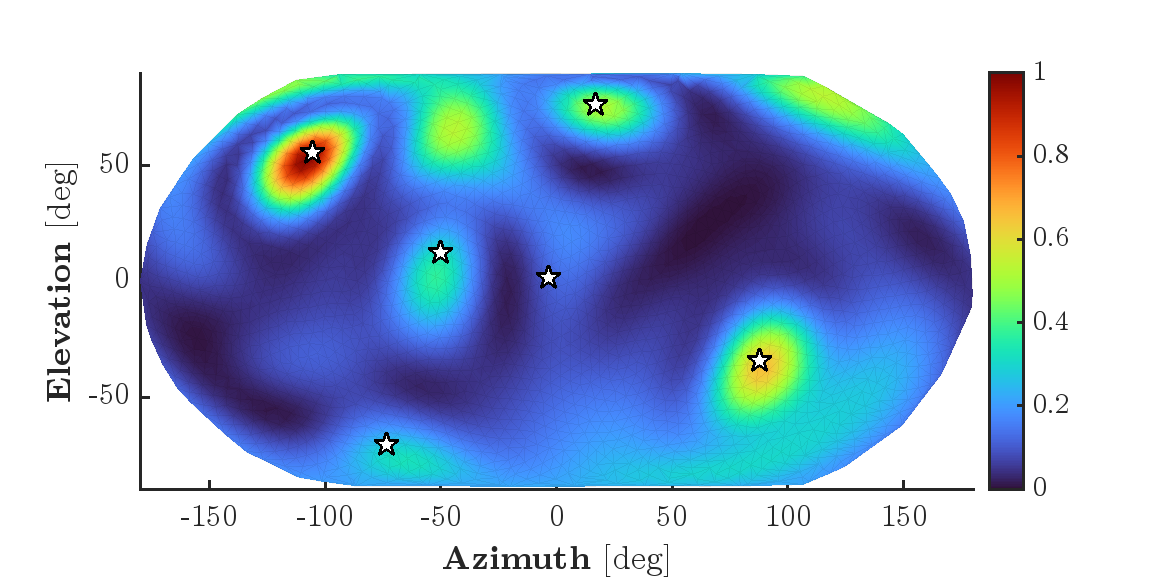}\label{fig:Sim6_MaxWDI}}%
    \subfigure[MUSIC]{\includegraphics[width=0.5\textwidth]{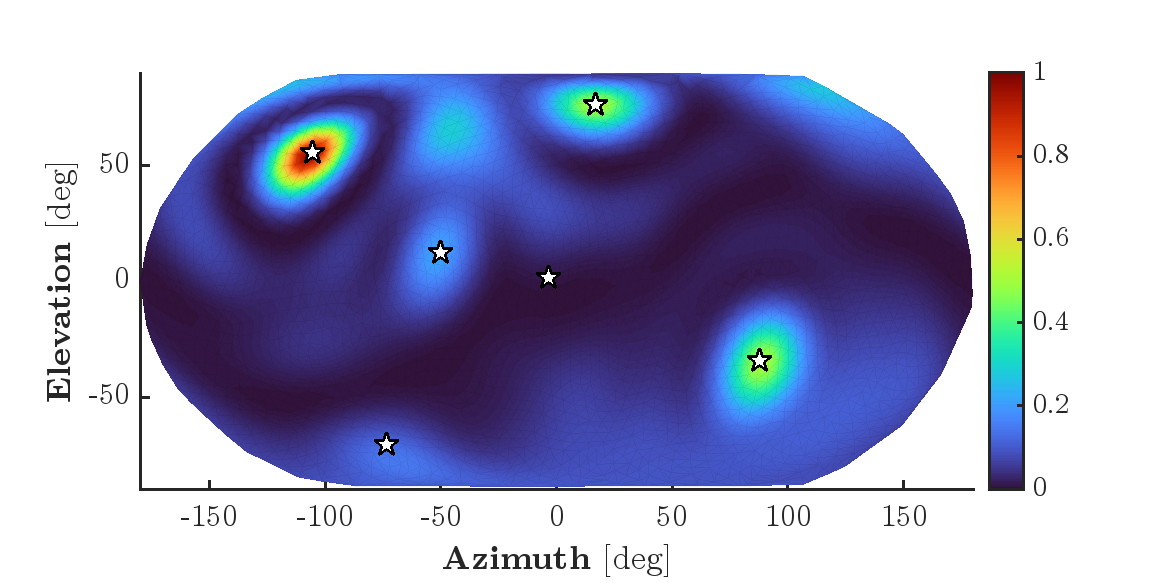}\label{fig:Sim6_MUSIC}}
  \caption{Localization function $\overline{\mathbf{m}}$, calculated following Eq.~(\ref{eq:hergLocFunc}), for a simulated frame containing $P_\mathrm{sim}=6$ echoes placed at $\vOmega_1=[-120^\circ,45.0^\circ]$, $\vOmega_2=[22.5^\circ,60.0^\circ]$, $\vOmega_3=[90.0^\circ,-22.5^\circ]$, and $\vOmega_4=[-87.0^\circ,-60.0^\circ]$, $\vOmega_1=[-55.0^\circ,12.0^\circ]$, $\vOmega_1=[0^\circ,0^\circ]$ respectively. The second through sixth echoes are synthesized at $-2.5$~dB, $-4.5$~dB, $-7$~dB, $-10.5$~dB and $-16.5$~dB relative to the first. White stars indicate the true simulated DoAs. The sphere is shown using the Equal Earth projection~\cite{savric_equal_2019}.}
  \label{fig:Sim6Echoes}
\end{figure*}

 Fig.~\ref{fig:Sim6Echoes} compares the localization functions obtained with the four methods considered in Sec.~\ref{sec:testSignals} for this six-echo frame. As with the EM32 case, the Herglotz localization function yields sharper, better-separated peaks at the true echo directions (white stars) than the SRP, maxWDI, and MUSIC alternatives, which increasingly merge or misplace the weaker reflections as the number of concurrent echoes grows.

\par Table~\ref{tab:peakDetSimRes} reports the corresponding peak localization accuracy. Consistent with the EM32 results of Sec.~\ref{sec:echoDetectEval}, the Herglotz-based localization function achieves the lowest angular error for four of the five reflections it detects, together with the smallest relative presence errors overall, and the highest detection rate ($83\%$) among the four methods. The MUSIC pseudospectrum, whose signal-subspace dimension is again constrained by eigenspectrum analysis, fails to resolve the weaker reflections and yields the lowest detection rate ($50\%$), while SRP and maxWDI fall between these extremes, with maxWDI over-segmenting the localization map into spurious peaks (detection rate above $100\%$).

\subsection{\label{sec:largeScaleSim}Large-Scale Statistical Simulation}

\par To evaluate the proposed method beyond individual example frames, a large-scale statistical simulation was conducted. Individual analysis frames containing between one and twelve reflections were simulated as described in Sec.~\ref{sec:testSignals}, using the EM64 geometry. For each number of reflections $P_\mathrm{sim}\in\{1,\dots,12\}$, $250$ independent scenarios were generated with randomly drawn incident directions, subject to a minimum angular separation of $35^\circ$ between echoes to avoid ambiguity from overlapping reflections. All reflections within a given frame were simulated with equal amplitude, isolating the influence of the number of concurrent echoes from that of their relative levels.

\par
\begin{figure*}[p]
  \centering
  \subfigure[Herglotz]{\includegraphics[width=0.5\textwidth]{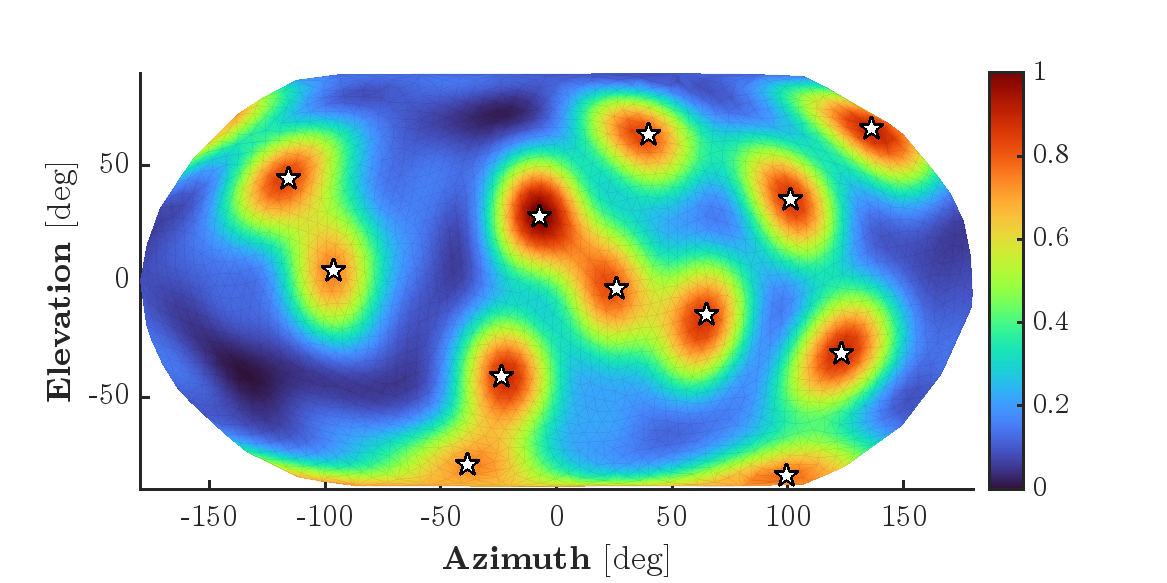}\label{fig:Sim12_herglotz}}%
  \subfigure[SRP]{\includegraphics[width=0.5\textwidth]{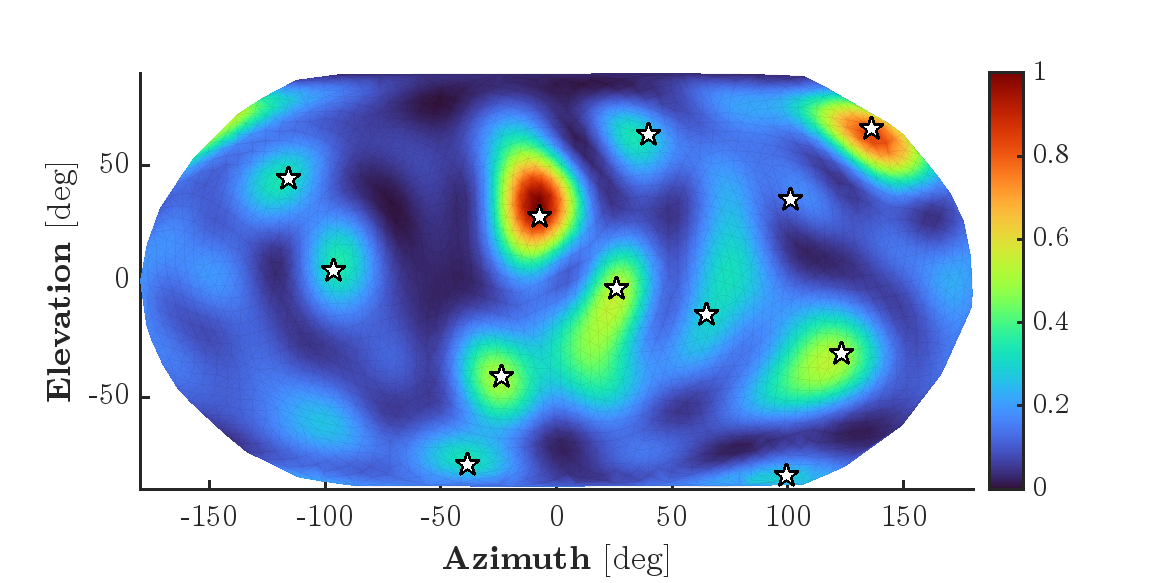}\label{fig:Sim12_SRP}}\\
  \subfigure[maxWDI]{\includegraphics[width=0.5\textwidth]{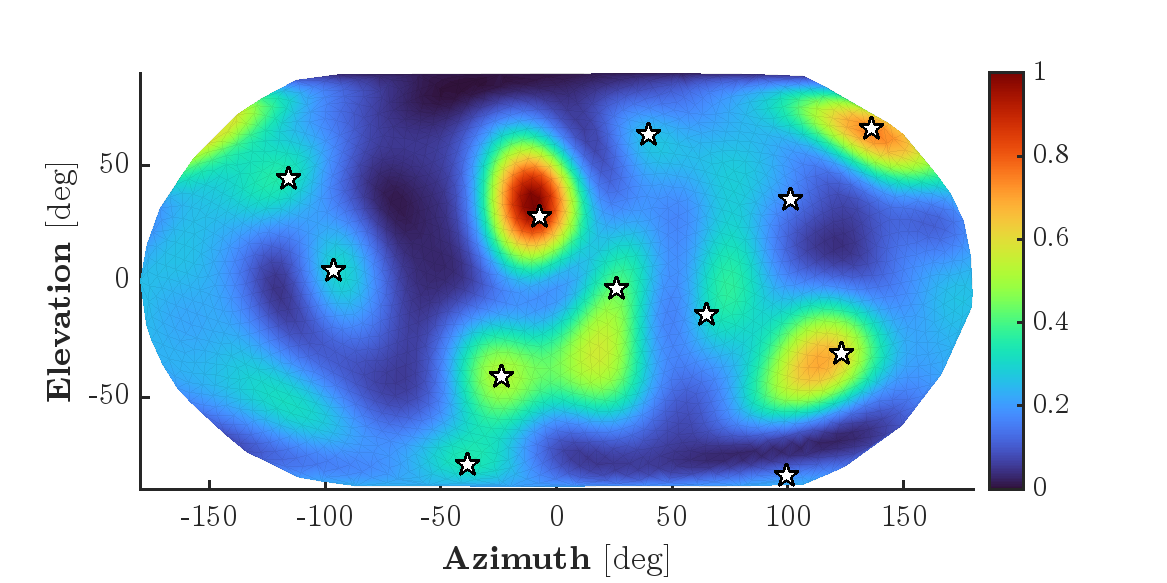}\label{fig:Sim12_MaxWDI}}%
  \subfigure[MUSIC]{\includegraphics[width=0.5\textwidth]{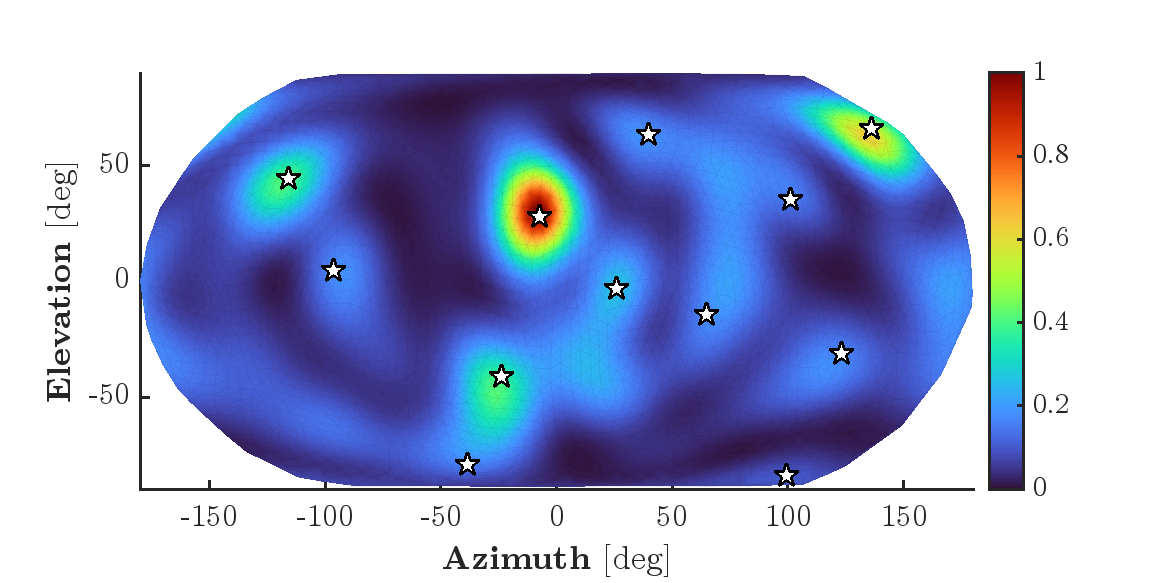}\label{fig:Sim12_MUSIC}}
  \caption{Localization function $\overline{\mathbf{m}}$, calculated following Eq.~(\ref{eq:hergLocFunc}), for a simulated frame containing $P_\mathrm{sim}=12$ echoes randomly placed with a minimum angular separation of $35^\circ$. All twelve impulses share the same gain. The sphere is shown using the Equal Earth projection~\cite{savric_equal_2019}.}
  \label{fig:Sim12Echoes}
\end{figure*}

 Fig.~\ref{fig:Sim12Echoes} illustrates the localization functions obtained with the four methods for a representative scenario containing $P_\mathrm{sim}=12$ randomly placed, equal-level echoes. As in the lower-order cases, the Herglotz localization function preserves twelve visually distinguishable peaks, whereas the reference methods exhibit greater smearing and spurious structure at this echo density.

\par
\begin{figure}[!h]%
\centering%
\includegraphics[width=\linewidth]{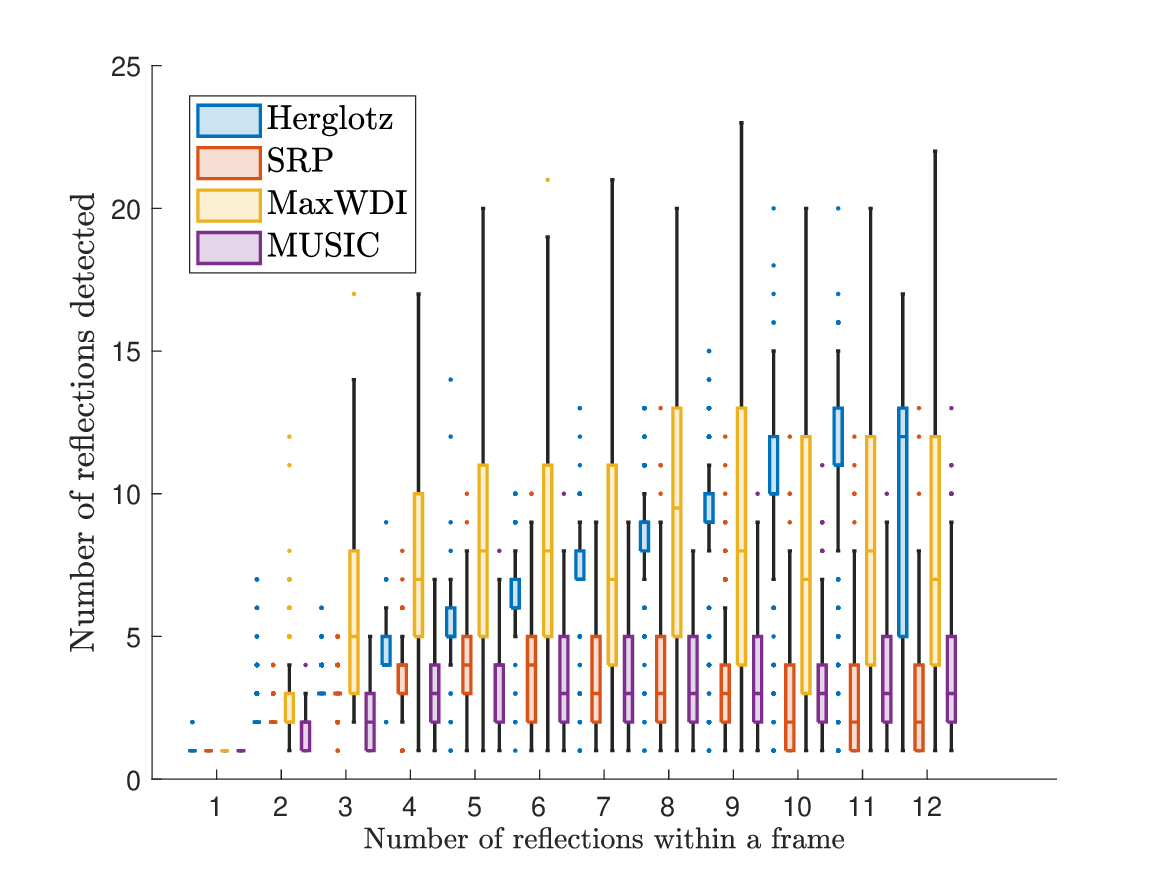}%
\caption{Number of echoes detected by the iterative Gaussian algorithm orf Sec.~\ref{sec:echoNumEst}, evaluated across $250$ DoA configurations for each number of simulated reflections, using the four localization map methods.}%
\label{fig:EchoedDetected}%
\end{figure}

\begin{figure*}[!h]
\centering
\subfigure[Angular Error]{\includegraphics[width=0.5\textwidth]{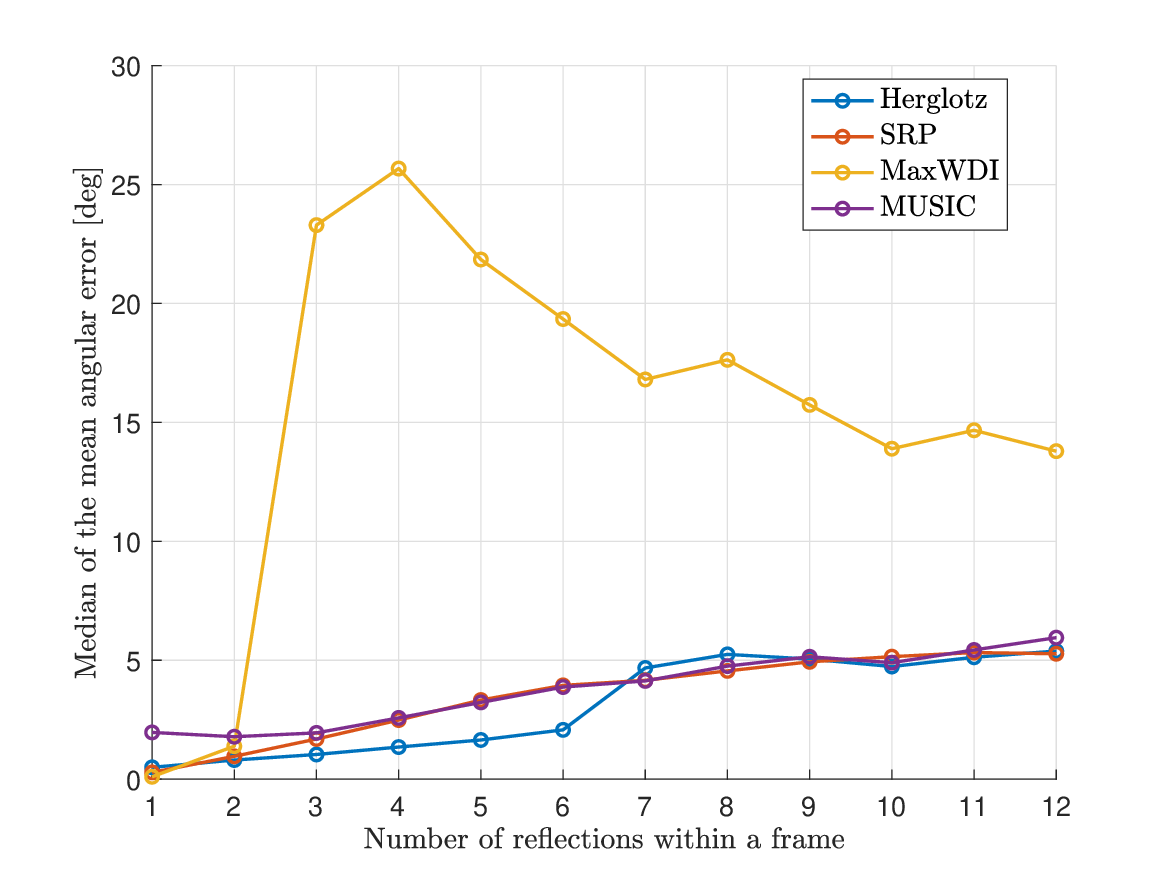}\label{fig:ErrAng}}%
\subfigure[Relative Presence Error]{\includegraphics[width=0.5\textwidth]{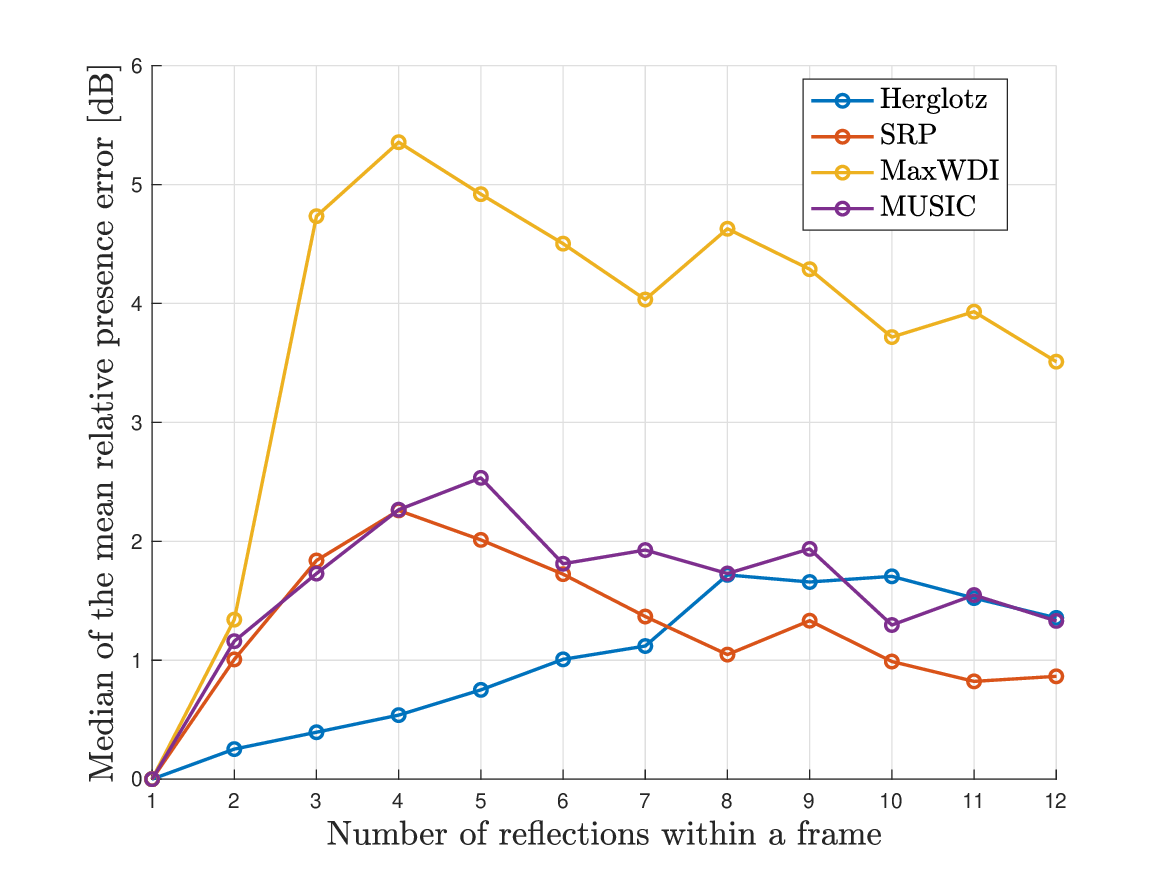}\label{fig:ErrPres}}\\
\caption{Medians of the mean angular error and relative presence error for detected reflections within a frame, evaluated across $250$ DoA configurations for each number of simulated reflections, using the four localization map methods.}
\label{fig:Errors big sim}
\end{figure*}

 Fig.~\ref{fig:EchoedDetected} summarizes, for each simulated number of reflections, the distribution of the number of echoes detected by the iterative Gaussian model-selection algorithm of Sec.~\ref{sec:echoNumEst}, applied to each of the four localization functions across the $250$ configurations. Fig.~\ref{fig:Errors big sim} shows the corresponding medians of the mean angular error and relative presence error, computed over the correctly matched, detected reflections in each frame. Notably, the Herglotz-based localization function detects substantially more reflections than SRP and MUSIC, particularly as the number of reflections increases. The resulting median errors at these higher reflection counts are therefore especially indicative of the method's ability to localize a larger number of concurrent reflections while maintaining accurate parameter estimates.

\subsection{\label{sec:espro}Measurement Validation}

\par The proposed method was further validated using measured SRIRs acquired in the ``Espace de projection'' (ESPRO), IRCAM's variable-acoustics concert hall \cite{Peutz:1978}. ESPRO is a shoebox-shaped room whose electroacoustic architecture allows continuous adjustment of the room's reverberant characteristics \cite{Noisternig:2013tv, Carpentier:2016}. The measurements analyzed here were collected as part of a larger measurement campaign in this space. The loudspeaker setup used during the campaign comprised a 75-loudspeaker HOA dome (Amadeus PMX-5 5-inch co-axial speakers), a nine-unit overhead speaker ring (d\&b E12 12-inch speakers), and four additional loudspeakers (d\&b E12 12-inch speakers) distributed on the floor. The microphone setup comprised ten ceiling-facing cardioid microphones (Neumann KM184), a binaural dummy head (Neumann KU100), and SMAs (mh-acoustics EM32 and EM64).

\par Two complementary evaluations were carried out on the measured EM64 SRIRs: a short-frame comparison focused on direct sound and early reflection localization accuracy, and a longer-frame comparison assessing the detection of a denser reflection sequence.

\subsubsection{Direct Sound and First Reflection}

\par
\begin{figure*}[p]
  \centering
    \subfigure[Herglotz]{\includegraphics[width=0.5\textwidth]{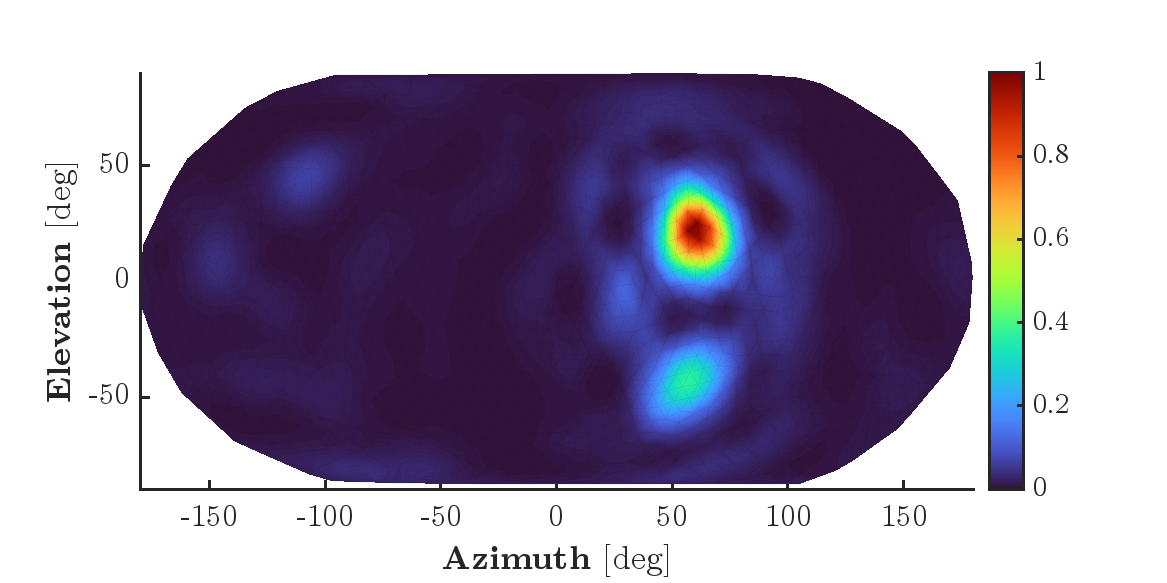}\label{fig:MeasureValid_herglotz}}%
    \subfigure[SRP]{\includegraphics[width=0.5\textwidth]{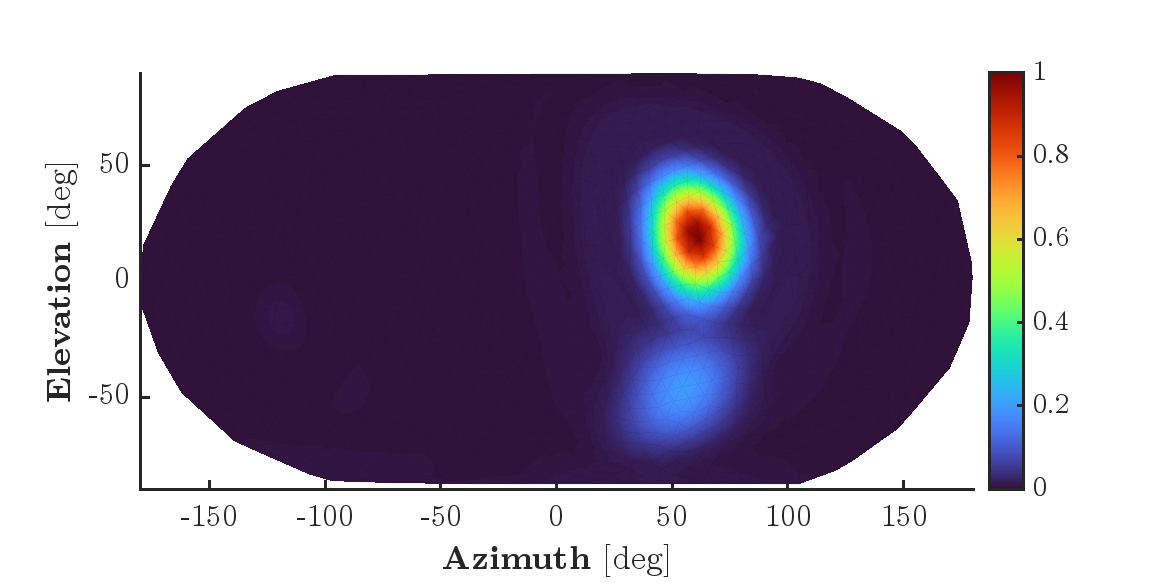}\label{fig:MeasureValid_origSRP}}\\
    \subfigure[maxWDI]{\includegraphics[width=0.5\textwidth]{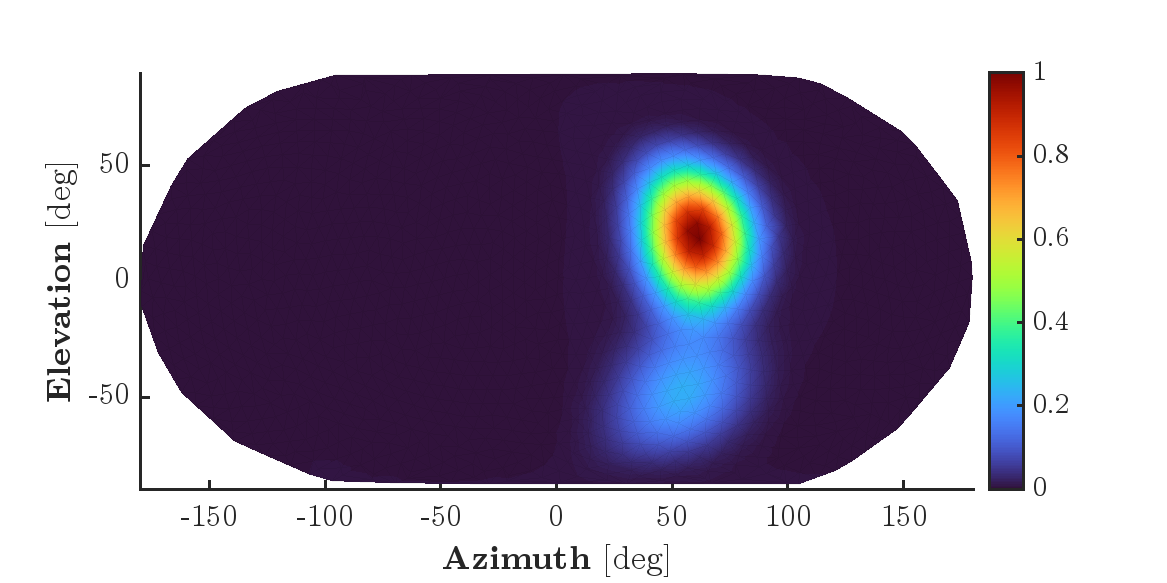}\label{fig:MeasureValid_SRPmaxWDI}}%
    \subfigure[MUSIC]{\includegraphics[width=0.5\textwidth]{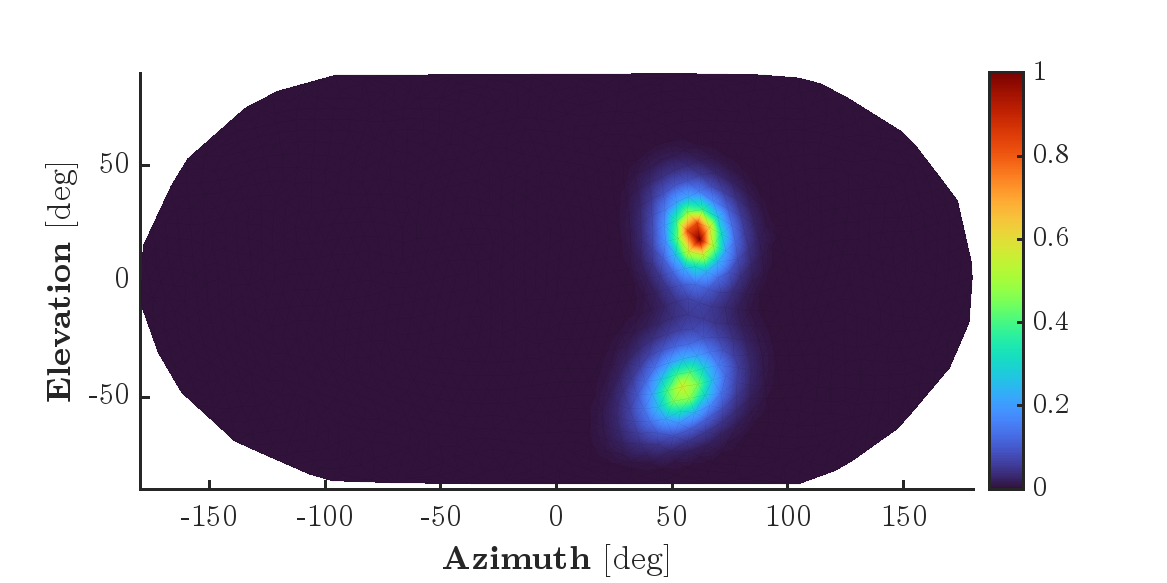}\label{fig:MeasureValid_MUSIC}}
  \caption{Localization function $\overline{\mathbf{m}}$, calculated following Eq.~(\ref{eq:hergLocFunc}), from a measured EM64 SRIR acquired in ESPRO. The visualization corresponds to a $10$~ms frame that includes both the direct sound and the first reflection. The sphere is shown using the Equal Earth projection~\cite{savric_equal_2019}.}
  \label{fig:MeasureValid}
\end{figure*}

 Fig.~\ref{fig:MeasureValid} shows the localization functions obtained with the four methods for a $10$~ms analysis frame containing the direct sound and the first reflection of a measured EM64 SRIR. Table~\ref{tab:peakDetMeasureRes} reports the corresponding angular and relative presence errors, where the reference (``true'') DoAs were obtained from an Image Source Method (ISM) simulation using the known loudspeaker and microphone positions.

\begin{table*}[!h]
  \caption{Peak Localization Accuracy Measurement\label{tab:peakDetMeasureRes}}
  \centering
  \begin{tabular}{|l||c|c||c|c||c|}
    \hline
    \rule{0pt}{2.8ex} \textbf{Method} & \multicolumn{2}{c||}{\textbf{Angular Errors} [deg]} & \multicolumn{2}{c||}{\textbf{Relative Presence Errors} [dB]} & \textbf{Detection Rate} [$\%$]\\
    \hline
    \hline
    \rule{0pt}{2.8ex} Herglotz & $1.5972$ & $\mathbf{1.0169}$  & 0 & $\mathbf{0.719}$ & 100\\
    \hline
    \rule{0pt}{2.8ex} SRP & $1.9586$ & $3.6250$ & 0 & $3.436$ & 100\\
    \hline
    \rule{0pt}{2.8ex} maxWDI & $2.4873$ & $6.2975$ & 0 & $3.614$ & 100\\
    \hline
    \rule{0pt}{2.8ex} \mbox{MUSIC} & $\mathbf{0.7771}$ & $2.8962$ & 0 & $5.752$ &  100\\
    \hline
  \end{tabular}
\end{table*}

\par All four methods correctly detect and localize the DoA of the direct sound and the first reflection, confirming that, for early, well-separated arrivals, the choice of localization function has limited impact on detection reliability. \mbox{MUSIC} achieves the lowest angular error for the direct sound, while the Herglotz method achieves the lowest angular error for the first reflection and the lowest relative presence error for both arrivals.

\subsubsection{Denser Reflection Sequence}

\par
\begin{figure*}[p]
  \centering
    \subfigure[Herglotz]{\includegraphics[width=0.5\textwidth]{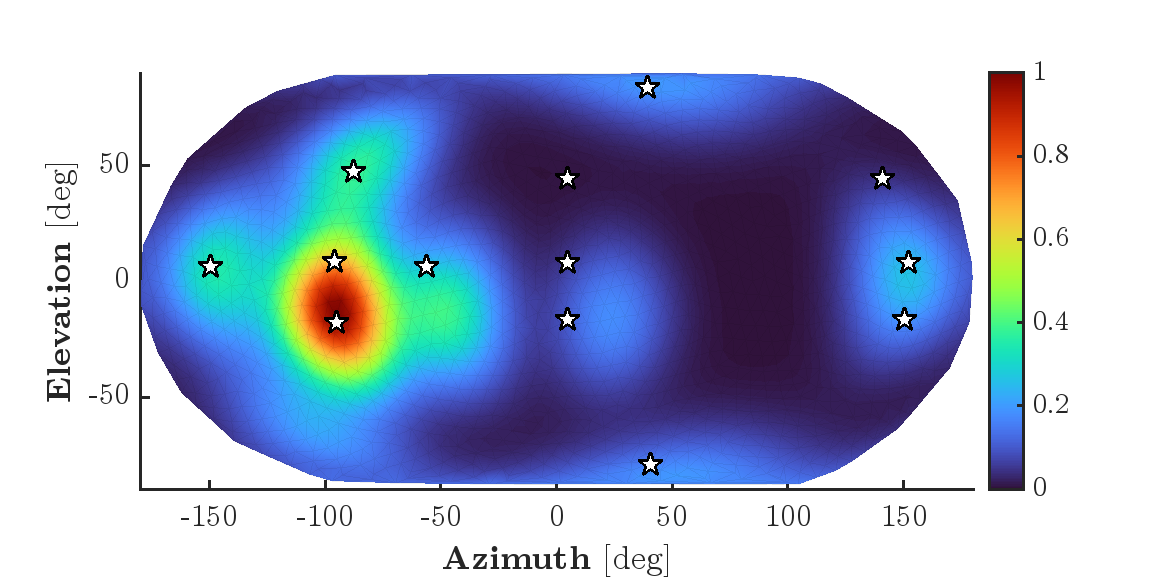}\label{fig:MeasureValidDetect_herglotz}}%
    \subfigure[SRP]{\includegraphics[width=0.5\textwidth]{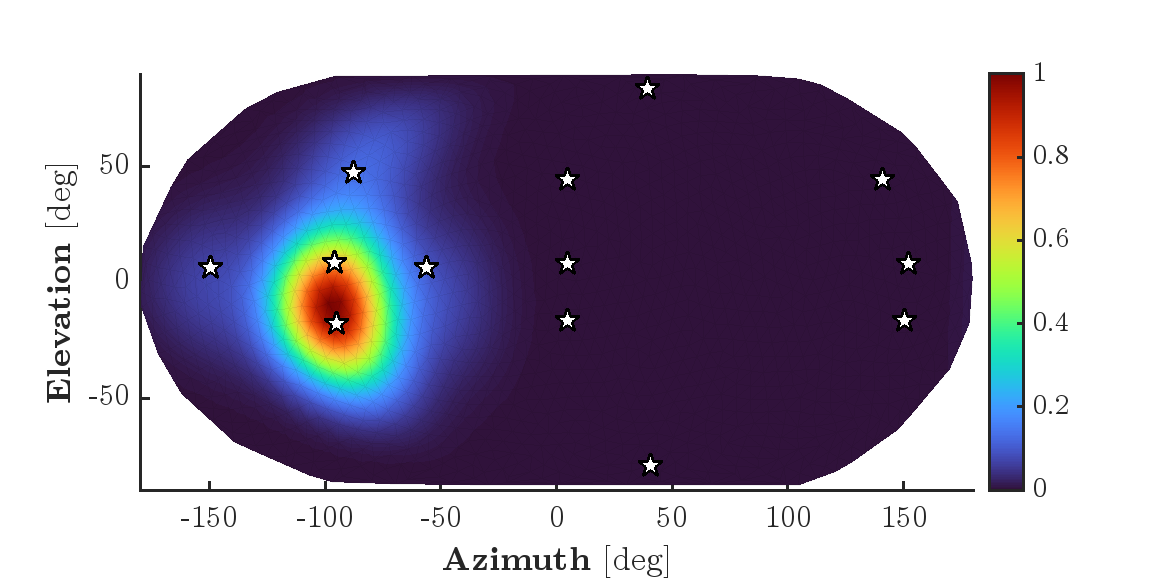}\label{fig:MeasureValidDetect_origSRP}}\\
    \subfigure[maxWDI]{\includegraphics[width=0.5\textwidth]{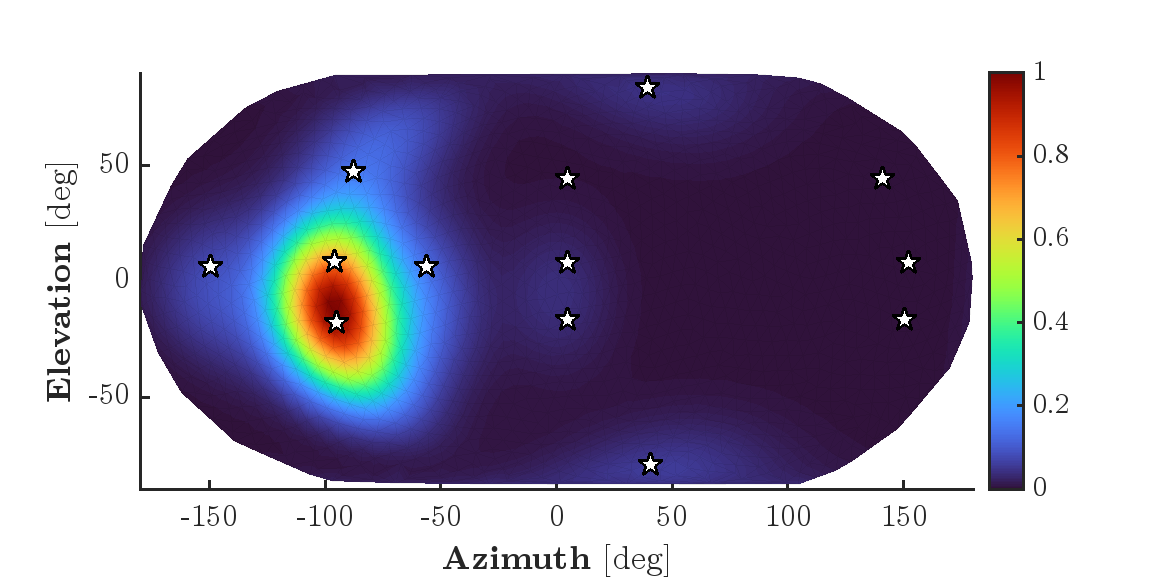}\label{fig:MeasureValidDetect_SRPmaxWDI}}%
    \subfigure[MUSIC]{\includegraphics[width=0.5\textwidth]{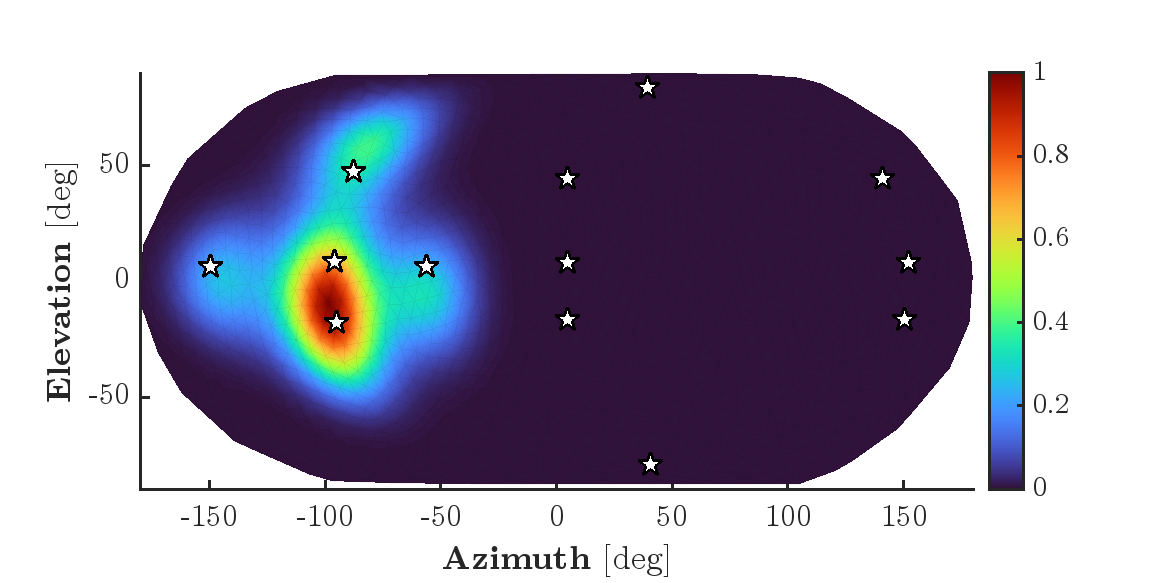}\label{fig:MeasureValidDetect_MUSIC}}
  \caption{Reconstructed localization function $\widehat{\mathbf{m}}\left(\mathbf{\Psi}\right)$ obtained from the radial Gaussian model of Eqs.~(\ref{eq:locFuncModel}) and~(\ref{eq:matLocFuncModel}) applied to the four localization functions, computed from a measured EM64 SRIR acquired in ESPRO. The visualization corresponds to a $42$~ms frame excluding the direct sound and the first two reflections, and incorporating the following $13$ reflections predicted by second-order ISM. White stars indicate the ISM-predicted reflection DoAs. The sphere is shown using the Equal Earth projection~\cite{savric_equal_2019}.}
  \label{fig:MeasureValidDetect}
\end{figure*}

 To assess performance under more realistic, denser reflection conditions, the analysis was extended to a $42$~ms frame excluding the direct sound and first two reflections but incorporating the following $13$ reflections, whose reference DoAs were computed using second-order ISM. Fig.~\ref{fig:MeasureValidDetect} shows the reconstructed localization maps obtained after applying the Gaussian detection algorithm of Sec.~\ref{sec:echoNumEst} to each of the four localization functions, with white stars marking the ISM-predicted reflection DoAs. Table~\ref{tab:EchoDetMeasureRes} summarizes the results.

\begin{table}[!h]
    \caption{Echo Detection Accuracy, ESPRO Measurement ($13$-Reflection Frame) \label{tab:EchoDetMeasureRes}}
    \centering
    \begin{tabular}{|l|c|}
        \hline
        \rule{0pt}{2.8ex} \textbf{Method} & \textbf{Detection Rate} [\%]\\
        \hline
        \hline
        \rule{0pt}{2.8ex} Herglotz & \textbf{70}\\
        \hline
        \rule{0pt}{2.8ex} SRP & 31\\
        \hline
        \rule{0pt}{2.8ex} maxWDI & 62\\
        \hline
        \rule{0pt}{2.8ex} \mbox{MUSIC} & 38\\
        \hline
    \end{tabular}
\end{table}

\par In this denser, more realistic reflection sequence, the Herglotz localization function substantially outperforms the reference methods, achieving a detection rate of $70\%$ against $31$--$62\%$ for SRP, maxWDI, and MUSIC. This measured result is consistent with the trends observed in the large-scale simulation of Sec.~\ref{sec:largeScaleSim}: as the number of concurrent, closely-spaced reflections within a single analysis frame grows, the proposed Herglotz formulation increasingly outperforms conventional SRP and subspace-based localization functions.

\par Taken together, the results of Secs.~\ref{sec:higherOrderSMA}--\ref{sec:espro} indicate that, while all four localization methods perform comparably for isolated, well-separated arrivals such as the direct sound and first reflection, the proposed Herglotz localization function offers a marked and consistent advantage as the echo density increases, in both higher-order simulated conditions and real SRIR measurements acquired in a variable-acoustics concert hall. This advantage is observed across all three complementary metrics considered throughout this section, namely angular error, relative presence error, and detection rate, particularly once at least four echoes are present within the same analysis frame.

\section{Conclusion}

\par This paper introduced a Herglotz wavefunction-based framework for localizing multiple early reflections within a single analysis frame of an SRIR measured by an SMA. By representing the measured sound field as a superposition of Herglotz wavefunctions, the proposed formulation naturally leads to a localization function whose peaks correspond to the directions of arrival of the incident echoes. A regularized inversion in the spherical harmonic domain was developed to estimate this localization function, followed by a radial Gaussian fitting procedure and an iterative information criterion-based strategy to jointly estimate both the directions of arrival and the number of incident echoes.

\par Simulations and measurements demonstrated that the proposed localization function provides a more faithful representation of multiple simultaneous reflections than conventional SRP- and MUSIC-based localization maps. Combined with the proposed Gaussian fitting procedure, the method achieved more accurate direction estimates, lower relative amplitude (presence) errors, and more reliable estimation of the number of incident echoes, particularly in analysis frames containing several closely spaced reflections. These results suggest that the Herglotz formulation is well suited to the analysis of the early part of SRIRs, where multiple broadband impulsive arrivals coexist within the same time frame.

\par The present work focuses on the localization of reflections within individual analysis frames as a first step toward a complete spatio-temporal analysis of early reflections. Future work will investigate frame-to-frame echo tracking and the construction of complete early-reflection cartographies, along with a more extensive evaluation using both simulated and measured SRIRs. Further studies should also examine the influence of the selected information criterion, the dynamic range between reflections, and the maximum spherical harmonic order supported by the SMA on localization and detection performance.

\section*{Acknowledgments}
This work was funded in part by the RASPUTIN project (grant number ANR-18-CE38-0004) and by the PostGenAI@Paris project (grant number ANR-23-IACL-0007), supported by the French government through the France 2030 program and coordinated by the Sorbonne Cluster for Artificial Intelligence (SCAI) at Sorbonne Université. The authors also acknowledge the support of the École doctorale Informatique, Télécommunications, et Électronique (EDITE) at Sorbonne Université. The authors would additionally like to thank Thibaut Carpentier and Olivier Warusfel for their numerous contributions to this work.

\bibliographystyle{IEEEtran}
\bibliography{biblio.bib}

@article{savric_equal_2019,
	title = {The {Equal} {Earth} {Map} {Projection}},
	volume = {33},
	issn = {1365-8816, 1362-3087},
	commenturl = {https://www.tandfonline.com/doi/full/10.1080/13658816.2018.1504949},
	doi = {10.1080/13658816.2018.1504949},
	number = {3},
	urldate = {2022-05-02},
	journal = {Int. J. Geogr. Inf. Sci.},
	author = {Šavrič, Bojan and Patterson, Tom and Jenny, Bernhard},
	month = mar,
	year = {2019},
	pages = {454--465},
	url = {https://doi.org/10.1080/13658816.2018.1504949}
}

@article{zotter_energy-preserving_2012,
	title = {Energy-{Preserving} {Ambisonic} {Decoding}},
	volume = {98},
	doi = {10.3813/AAA.918490},
	number = {1},
	journal = {Acta Acust. united Ac.},
	author = {Zotter, Franz and Pomberger, Hannes and Noisternig, Markus},
	month = feb,
	year = {2012},
	pages = {37--47},
	url = {https://doi.org/10.3813/AAA.918490},
}

@article{akaike_new_1974,
	title = {A {New} {Look} at the {Statistical} {Model} {Identification}},
	volume = {19},
	issn = {0018-9286},
	commenturl = {http://ieeexplore.ieee.org/document/1100705/},
	doi = {10.1109/TAC.1974.1100705},
	language = {en},
	number = {6},
	urldate = {2022-04-21},
	journal = {IEEE Trans. Automat. Contr.},
	author = {Akaike, Hirotugu},
	month = dec,
	year = {1974},
	pages = {716--723},
	url = {https://doi.org/10.1109/TAC.1974.1100705},
}

@article{barzilai_two-point_1988,
	title = {Two-{Point} {Step} {Size} {Gradient} {Methods}},
	volume = {8},
	issn = {0272-4979, 1464-3642},
	commenturl = {https://academic.oup.com/imajna/article-lookup/doi/10.1093/imanum/8.1.141},
	doi = {10.1093/imanum/8.1.141},
	language = {en},
	number = {1},
	urldate = {2022-05-25},
	journal = {IMA J. Numer. Anal.},
	author = {Barzilai, Jonathan and Borwein, Jonathan M.},
	year = {1988},
	pages = {141--148},
	url = {https://doi.org/10.1093/imanum/8.1.141},
}

@incollection{hansen_regularization_1998,
	title = {Regularization {Tools}},
	isbn = {978-0-89871-403-6 978-0-89871-969-7},
	commenturl = {http://epubs.siam.org/doi/book/10.1137/1.9780898719697},
	language = {en},
	urldate = {2021-12-01},
	booktitle = {Rank-{Deficient} and {Discrete} {Ill}-{Posed} {Problems}: {Numerical} {Aspects} of {Linear} {Inversion}},
	publisher = {Soc. Ind. Appl. Math.},
	author = {Hansen, Per Christian},
	month = jan,
	year = {1998},
	doi = {10.1137/1.9780898719697},
	url = {https://doi.org/10.1137/1.9780898719697},
}

@book{colton_inverse_2013,
	address = {New York, NY},
	series = {Applied {Mathematical} {Sciences}},
	title = {Inverse {Acoustic} and {Electromagnetic} {Scattering} {Theory}},
	volume = {93},
	isbn = {978-1-4614-4941-6 978-1-4614-4942-3},
	commenturl = {http://link.springer.com/10.1007/978-1-4614-4942-3},
	language = {en},
	urldate = {2022-02-16},
	publisher = {Springer New York},
	author = {Colton, David and Kress, Rainer},
	year = {2013},
	doi = {10.1007/978-1-4614-4942-3},
	url = {https://doi.org/10.1007/978-1-4614-4942-3},
}

@article{rafaely_plane-wave_2004,
	title = {Plane-{Wave} {Decomposition} of the {Sound} {Field} on a {Sphere} by {Spherical} {Convolution}},
	volume = {116},
	issn = {0001-4966},
	commenturl = {http://asa.scitation.org/doi/10.1121/1.1792643},
	doi = {10.1121/1.1792643},
	language = {en},
	number = {4},
	urldate = {2022-03-02},
	journal = {J. Acoust. Soc. Am.},
	author = {Rafaely, Boaz},
	month = oct,
	year = {2004},
	pages = {2149--2157},
	url = {https://doi.org/10.1121/1.1792643},
}

@article{li_flexible_2007,
	title = {Flexible and {Optimal} {Design} of {Spherical} {Microphone} {Arrays} for {Beamforming}},
	volume = {15},
	issn = {1558-7916},
	commenturl = {http://ieeexplore.ieee.org/document/4067056/},
	doi = {10.1109/TASL.2006.876764},
	number = {2},
	urldate = {2022-03-07},
	journal = {IEEE Trans. Audio Speech Lang. Process.},
	author = {Li, Zhiyun and Duraiswami, Ramani},
	month = feb,
	year = {2007},
	pages = {702--714},
	url = {https://doi.org/10.1109/TASL.2006.876764},
}

@phdthesis{daniel_representation_2001,
	address = {Paris, France},
	title = {Représentation de champs acoustiques, application à la transmission et à la reproduction de scènes sonores complexes dans un contexte multimédia},
	language = {en},
	school = {Université de Paris 6},
	author = {Daniel, Jérôme},
	year = {2001},
}

@inbook{noisternig_reconstructing_2011,
	address = {Singapore},
	title = {Reconstructing {Sound} {Source} {Directivity} in {Virtual} {Acoustic} {Environments}},
	booktitle = {Principles and {Applications} of {Spatial} {Hearing}},
	publisher = {World Scientific Publishing},
	author = {Noisternig, Markus and Zotter, Franz and Katz, Brian F. G.},
	editor = {Suzuki, Yôiti and Brungart, Douglas and Iwaya, Yukio and Iida, Kazuhiro and Cabrera, Densil and Kato, Hiroaki},
	doi = {10.1142/9789814299312_0028},
	commenturl = {https://www.worldscientific.com/doi/abs/10.1142/9789814299312_0028},
	pages = {357--372},
	year = {2011},
	url = {https://doi.org/10.1142/9789814299312_0028},
}

@inproceedings{rafaely_spatial_2008,
	address = {Trento, Italy},
	title = {Spatial {Sampling} and {Beamforming} for {Spherical} {Microphone} {Arrays}},
	isbn = {978-1-4244-2337-8},
	commenturl = {http://ieeexplore.ieee.org/document/4538673/},
	doi = {10.1109/HSCMA.2008.4538673},
	urldate = {2022-03-02},
	booktitle = {Proc. {Hands}-{Free} {Speech} {Communication} and {Microphone} {Arrays}},
	commentpublisher = {IEEE},
	author = {Rafaely, Boaz},
	month = may,
	year = {2008},
	pages = {5--8},
	url = {https://doi.org/10.1109/HSCMA.2008.4538673},
}

@article{rafaely_analysis_2005,
	title = {Analysis and {Design} of {Spherical} {Microphone} {Arrays}},
	volume = {13},
	issn = {1063-6676},
	commenturl = {http://ieeexplore.ieee.org/document/1369318/},
	doi = {10.1109/TSA.2004.839244},
	number = {1},
	urldate = {2022-03-02},
	journal = {IEEE Trans. Speech Audio Process.},
	author = {Rafaely, B.},
	month = jan,
	year = {2005},
	pages = {135--143},
	url = {https://doi.org/10.1109/TSA.2004.839244},
}

@inproceedings{abhayapala_theory_2002,
	address = {Orlando, U.S.A.},
	title = {Theory and {Design} of {High} {Order} {Sound} {Field} {Microphones} {Using} {Spherical} {Microphone} {Array}},
	booktitle = {Proc. {IEEE} {Int.} {Conf.} on {Acoust.}, {Speech}, and {Signal} {Proc.} ({ICASSP})},
	author = {Abhayapala, Thushara D. and Ward, Darren B.},
	year = {2002},
	volume = {2},
	pages = {II-1949-II-1952},
	doi = {10.1109/ICASSP.2002.5745011},
	url = {https://doi.org/10.1109/ICASSP.2002.5745011},
}

@inproceedings{meyer_highly_2002,
	address = {Orlando, U.S.A.},
	title = {A {Highly} {Scalable} {Spherical} {Microphone} {Array} {Based} {On} an {Orthonormal} {Decomposition} of the {Soundfield}},
	booktitle = {Proc. {IEEE} {Int.} {Conf.} on {Acoust.}, {Speech}, and {Signal} {Proc.} ({ICASSP})},
	author = {Meyer, Jens and Elko, Gary},
	year = {2002},
	 volume={2},
	pages = {II-1781-II-1784},
}

@inproceedings{daniel_further_2004,
	address = {Berlin, Germany},
	title = {Further {Study} of {Sound} {Field} {Coding} with {Higher} {Order} {Ambisonics}},
	booktitle = {Proc 116th {Audio} {Eng.} {Soc.} {Conv.}},
	author = {Daniel, Jérôme and Moreau, Sébastien},
	year = {2004},
	pages = {14},
}

@phdthesis{masse_analysis_2022,
	address = {Paris, France},
	title = {Analysis, {Treatment}, and {Manipulation} {Methods} for {Spatial} {Room} {Impulse} {Responses} {Measured} with {Spherical} {Microphone} {Arrays}},
	school = {Sorbonne Université},
	author = {Massé, Pierre},
	month = feb,
	year = {2022},
}

@article{gotz_mixing_2015,
	title = {Mixing {Time} {Prediction} using {Spherical} {Microphone} {Arrays}},
	volume = {137},
	issn = {0001-4966},
	commenturl = {http://asa.scitation.org/doi/10.1121/1.4907547},
	doi = {10.1121/1.4907547},
	number = {2},
	urldate = {2022-04-19},
	journal = {J. Acoust. Soc. Am.},
	author = {Götz, Philipp and Kowalczyk, Konrad and Silzle, Andreas and Habets, Emanuël A. P.},
	month = feb,
	year = {2015},
	pages = {EL206--EL212},
	url = {https://doi.org/10.1121/1.4907547},
}

@article{polack_playing_1993,
	title = {Playing {Billiards} in the {Concert} {Hall}: {The} {Mathematical} {Foundations} of {Geometrical} {Room} {Acoustics}},
	volume = {38},
	issn = {0003682X},
	shorttitle = {Playing billiards in the concert hall},
	commenturl = {https://linkinghub.elsevier.com/retrieve/pii/0003682X9390054A},
	doi = {10.1016/0003-682X(93)90054-A},
	number = {2-4},
	urldate = {2022-04-19},
	journal = {Appl. Acoust.},
	author = {Polack, Jean-Dominique},
	year = {1993},
	pages = {235--244},
	url = {https://doi.org/10.1016/0003-682X(93)90054-A},
}

@article{schroeder_new_1965,
	title = {New {Method} of {Measuring} {Reverberation} {Time}},
	volume = {37},
	language = {en},
	number = {3},
	journal = {J. Acoust. Soc. Am.},
	author = {Schroeder, Manfred Robert},
	year = {1965},
	pages = {409--412},
	doi = {10.1121/1.1909343},
	url = {https://doi.org/10.1121/1.1909343},
}

@article{morgenstern_modal_2018,
	title = {Modal {Smoothing} for {Analysis} of {Room} {Reflections} {Measured} with {Spherical} {Microphone} and {Loudspeaker} {Arrays}},
	volume = {143},
	issn = {0001-4966},
	commenturl = {http://asa.scitation.org/doi/10.1121/1.5024234},
	doi = {10.1121/1.5024234},
	language = {en},
	number = {2},
	urldate = {2022-04-21},
	journal = {J. Acoust. Soc. Am.},
	author = {Morgenstern, Hai and Rafaely, Boaz},
	month = feb,
	year = {2018},
	pages = {1008--1018},
	url = {https://doi.org/10.1121/1.5024234},
}

@article{morgenstern_mimo_2017, 
	title = {{Design framework for spherical microphone and loudspeaker arrays in a multiple-input multiple-output system}},
	volume = {141},
	issn = {0001-4966}, 
	year = {2017}, 
	author = {Morgenstern, Hai and Rafaely, Boaz and Noisternig, Markus}, 
	journal = {J. Acoust. Soc. Am.}, 
	doi = {10.1121/1.4978660}, 
	url = {http://asa.scitation.org/doi/10.1121/1.4978660}, 
	pages = {2024 -- 2038}, 
	number = {3},
	language = {en}, 
	note = {doi: 10.1121/1.4978660},  
	month = {03}
}

@article{huleihel_spherical_2013,
	title = {Spherical {Array} {Processing} for {Acoustic} {Analysis} using {Room} {Impulse} {Responses} and {Time}-{Domain} {Smoothing}},
	volume = {133},
	issn = {0001-4966},
	commenturl = {http://asa.scitation.org/doi/10.1121/1.4804314},
	doi = {10.1121/1.4804314},
	language = {en},
	number = {6},
	urldate = {2022-04-21},
	journal = {J. Acoust. Soc. Am.},
	author = {Huleihel, Nejem and Rafaely, Boaz},
	month = jun,
	year = {2013},
	pages = {3995--4007},
	url = {https://doi.org/10.1121/1.4804314},
}

@article{khaykin_acoustic_2012,
	title = {Acoustic {Analysis} by {Spherical} {Microphone} {Array} {Processing} of {Room} {Impulse} {Responses}},
	volume = {132},
	issn = {0001-4966},
	commenturl = {http://asa.scitation.org/doi/10.1121/1.4726012},
	doi = {10.1121/1.4726012},
	language = {en},
	number = {1},
	urldate = {2022-04-21},
	journal = {J. Acoust. Soc. Am.},
	author = {Khaykin, Dima and Rafaely, Boaz},
	month = jul,
	year = {2012},
	pages = {261--270},
	url = {https://doi.org/10.1121/1.4726012},
}

@article{roy_esprit_1989,
	title = {{ESPRIT} – {Estimation} of {Signal} {Parameters} via {Rotational} {Invariance} {Techniques}},
	volume = {37},
	issn = {00963518},
	commenturl = {http://ieeexplore.ieee.org/document/32276/},
	doi = {10.1109/29.32276},
	number = {7},
	urldate = {2022-04-21},
	journal = {IEEE Trans. Audio Speech Lang. Process.},
	author = {Roy, Richard and Kailath, Thomas},
	month = jul,
	year = {1989},
	pages = {984--995},
	url = {https://doi.org/10.1109/29.32276},
}

@article{schmidt_multiple_1986,
	title = {Multiple {Emitter} {Location} and {Signal} {Parameter} {Estimation}},
	volume = {34},
	issn = {0096-1973},
	commenturl = {http://ieeexplore.ieee.org/document/1143830/},
	doi = {10.1109/TAP.1986.1143830},
	language = {en},
	number = {3},
	urldate = {2022-04-21},
	journal = {IEEE Trans. Antennas Propagat.},
	author = {Schmidt, Ralph O.},
	month = mar,
	year = {1986},
	pages = {276--280},
	url = {https://doi.org/10.1109/TAP.1986.1143830},
}

@inproceedings{mccormack_sharpening_2019,
	address = {Brighton, U.K.},
	title = {Sharpening of {Angular} {Spectra} {Based} on a {Directional} {Re}-assignment {Approach} for {Ambisonic} {Sound}-{Field} {Visualisation}},
	isbn = {978-1-4799-8131-1},
	commenturl = {https://ieeexplore.ieee.org/document/8683621/},
	doi = {10.1109/ICASSP.2019.8683621},
	urldate = {2021-11-02},
	booktitle = {Proc. {IEEE} {Int.} {Conf.} on {Acoust.}, {Speech} and {Signal} {Proc.} ({ICASSP})},
	commentpublisher = {IEEE},
	author = {McCormack, Leo and Politis, Archontis and Pulkki, Ville},
	month = may,
	year = {2019},
	pages = {576--580},
	url = {https://doi.org/10.1109/ICASSP.2019.8683621},
}

@article{mccormack_higher-order_2020,
	title = {Higher-{Order} {Spatial} {Impulse} {Response} {Rendering}: {Investigating} the {Perceived} {Effects} of {Spherical} {Order}, {Dedicated} {Diffuse} {Rendering}, and {Frequency} {Resolution}},
	volume = {68},
	issn = {15494950},
	shorttitle = {Higher-{Order} {Spatial} {Impulse} {Response} {Rendering}},
	commenturl = {http://www.aes.org/e-lib/browse.cfm?elib=20852},
	doi = {10.17743/jaes.2020.0026},
	language = {en},
	number = {5},
	urldate = {2022-03-08},
	journal = {J. Audio Eng. Soc.},
	author = {McCormack, Leo and Pulkki, Ville and Politis, Archontis and Scheuregger, Oliver and Marschall, Marton},
	month = jun,
	year = {2020},
	pages = {338--354},
	url = {https://doi.org/10.17743/jaes.2020.0026},
}

@article{tervo_spatial_2013,
	title = {Spatial {Decomposition} {Method} for {Room} {Impulse} {Responses}},
	volume = {61},
	language = {en},
	number = {1},
	journal = {J. Audio Eng. Soc.},
	author = {Tervo, Sakari and Tynen, Jukka Pa and Kuusinen, Antti and Lokki, Tapio},
	year = {2013},
	pages = {12},
}

@article{merimaa_spatial_2005,
	title = {Spatial {Impulse} {Response} {Rendering} {I}: {Analysis} and {Synthesis}},
	volume = {53},
	language = {en},
	number = {12},
	journal = {J. Audio Eng. Soc.},
	author = {Merimaa, Juha and Pulkki, Ville},
	year = {2005},
	pages = {14},
}

@article{masse_denoising_2020,
	title = {Denoising {Directional} {Room} {Impulse} {Responses} with {Spatially} {Anisotropic} {Late} {Reverberation} {Tails}},
	volume = {10},
	issn = {2076-3417},
	commenturl = {https://www.mdpi.com/2076-3417/10/3/1033},
	doi = {10.3390/app10031033},
	number = {3},
	urldate = {2022-04-19},
	journal = {Appl. Sc.},
	author = {Massé, Pierre and Carpentier, Thibaut and Warusfel, Olivier and Noisternig, Markus},
	month = feb,
	year = {2020},
	pages = {1033},
	url = {https://doi.org/10.3390/app10031033},
}

@article{masse_robust_2020,
	title = {A {Robust} {Denoising} {Process} for {Spatial} {Room} {Impulse} {Responses} with {Diffuse} {Reverberation} {Tails}},
	volume = {147},
	issn = {0001-4966},
	commenturl = {http://asa.scitation.org/doi/10.1121/10.0001070},
	doi = {10.1121/10.0001070},
	language = {en},
	number = {4},
	urldate = {2022-04-19},
	journal = {J. Acoust. Soc. Am.},
	author = {Massé, Pierre and Carpentier, Thibaut and Warusfel, Olivier and Noisternig, Markus},
	month = apr,
	year = {2020},
	pages = {2250--2260},
	url = {https://doi.org/10.1121/10.0001070},
}

@inproceedings{carpentier_hybrid_2014,
	address = {Erlangen, Germany},
	title = {Hybrid {Reverberation} {Processor} with {Perceptual} {Control}},
	booktitle = {Proc 17th {Int.} {Conf.} on {Digital} {Audio} {Effects}},
	author = {Carpentier, Thibaut and Noisternig, Markus and Warusfel, Olivier},
	year = {2014},
}

@article{Chardon:2015hd,
    year = {2015},
    title = {{Design of Spatial Microphone Arrays for Sound Field Interpolation}},
    author = {Chardon, Gilles and Kreuzer, Wolfgang and Noisternig, Markus},
    journal = {IEEE J. Sel. Top. Signal Process.},
    doi = {10.1109/jstsp.2015.2412097},
    commenturl = {http://ieeexplore.ieee.org/xpl/articleDetails.jsp?tp=\&arnumber=7058409\&matchBoolean\%3Dtrue\%26rowsPerPage\%3D30\%26searchField\%3DSearch\_All\%26queryText\%3D\%28p\_Authors\%3Anoisternig\%29},
    pages = {780 -- 790},
    number = {5},
    volume = {9},
    month = {00},
    url = {https://doi.org/10.1109/jstsp.2015.2412097},
}

@article{schwarz_estimating_1978,
	title = {Estimating the {Dimension} of a {Model}},
	volume = {6},
	number = {2},
	journal = {The Annals of Statistics},
	author = {Schwarz, Gideon},
	year = {1978},
	pages = {461--464},
	url = {https://doi.org/10.1214/aos/1176344136}
}

@inproceedings{Peutz:1978,
	address = {Trento, Italy},
	title = {The {Variable} {Acoustics} of the `{Espace} de projection' of {IRCAM} (Paris)},
	urldate = {2025-03-02},
	booktitle = {Proc.\ 59th {AEES} {Convention}},
	commentpublisher = {aes},
	author = {Peutz, Victor M. A.},
	month = may,
	year = {1978},
	url = {www.aes.org/e-lib/browse.cfm?elib=3044}
}

@article{Carpentier:2016, 
	year = {2016}, 
	title = {{Holophonic Sound in IRCAM's Concert Hall: Technological and Aesthetic Practices}}, 
	author = {Carpentier, Thibaut and Barrett, Natasha and Gottfried, Rama and Noisternig, Markus}, 
	journal = {Computer Music Journal}, 
	issn = {0148-9267}, 
	doi = {10.1162/comj\_a\_00383}, 
	url = {http://www.mitpressjournals.org/doi/10.1162/COMJ\_a\_00383}, 
	pages = {14 -- 34}, 
	number = {4}, 
	volume = {40}, 
	month = {12}
}

@inproceedings{Noisternig:2013tv, 
	year = {2013}, 
	author = {Noisternig, Markus and Carpentier, Thibaut and Warusfel, Olivier}, 
	title = {{A Multichannel Loudspeaker Array for WFS/HOA Sound Spatialization at Ircam’s Concert Hall}},
	booktitle = {Proc.\ AIA-DAGA Conference on Acoustics},
	month = {03}
}

@ARTICLE{shlomo_blind_2021,
  author={Shlomo, Tom and Rafaely, Boaz},
  journal={IEEE Transactions on Signal Processing}, 
  title={Blind Localization of Early Room Reflections Using Phase Aligned Spatial Correlation}, 
  year={2021},
  volume={69},
  number={},
  pages={1213-1225},
  doi={10.1109/TSP.2021.3057495}
  }

@book{Williams:1999wm, 
	year = {1999}, 
	title = {{Fourier Acoustics: Sound Radiation and Nearfield Acoustical Holography}}, 
	author = {Williams, Earl G}, 
	series = {Academic Press}, 
	publisher = {Academic Press}
}

@ARTICLE{WarAbh01,
  author={Ward, D.B. and Abhayapala, T.D.},
  journal={IEEE Transactions on Speech and Audio Processing}, 
  title={Reproduction of a plane-wave sound field using an array of loudspeakers}, 
  year={2001},
  volume={9},
  number={6},
  pages={697--707},
  doi={10.1109/89.943347}
}

\appendices
\counterwithin{figure}{section}
\section{Wide Comparison Plots}

\vfill

\end{document}